\documentstyle[epsfig,psfig]{mn}

\newif\ifAMStwofonts

\def \z{\phantom{0}} 
\def \eg{e.g.,\/}
\def \ie{i.e.,\/}
\def \Hb{${\rmn H}\beta$}
\def \HgA{${\rmn H}\gamma_{\rmn A}$}
\def \HgF{${\rmn H}\gamma_{\rmn F}$}
\def \HgAF{${\rmn H}\gamma_{\rmn A, F}$}
\def \Hg{${\rmn H}\gamma$}

\def \HdF{${\rmn H}\delta_{\rmn F}$}
\def \HdAF{${\rmn H}\delta_{\rmn A, F}$}
\def \Hd{${\rmn H}\delta$}
\def \oii{[O{\small II}]$\lambda 3727$}
\def \oiia{[O{\small II}]}
\def \oiii{[O{\small III}]}
\def \oiiia{[O{\small III}]$\lambda 4959$}
\def \oiiib{[O{\small III}]$\lambda 5007$}
\def \fe{$<$Fe$>$}
\def \mgb{Mg{\,\it b}}

\def \kms{km\,s$^{-1}$}
\def \kmsM{km\,s$^{-1}$\,Mpc$^{-1}$}

\def \gon{Gonz\'{a}lez}
\def \hmpc{$h^{-1}$\,Mpc}

\ifoldfss
  \newcommand{\rmn}[1] {{\rm #1}}

  \ifCUPmtlplainloaded \else
    \NewTextAlphabet{textbfit} {cmbxti10} {}
    \NewTextAlphabet{textbfss} {cmssbx10} {}
    \NewMathAlphabet{mathbfit} {cmbxti10} {} 
    \NewMathAlphabet{mathbfss} {cmssbx10} {} 
  \fi
  \ifAMStwofonts
    \ifCUPmtlplainloaded \else
      \NewSymbolFont{upmath} {eurm10}
      \NewSymbolFont{AMSa} {msam10}
      \NewMathSymbol{\upi}     {0}{upmath}{19}
      \NewMathSymbol{\umu}     {0}{upmath}{16}
      \NewMathSymbol{\upartial}{0}{upmath}{40}
      \NewMathSymbol{\leqslant}{3}{AMSa}{36}
      \NewMathSymbol{\geqslant}{3}{AMSa}{3E}

      \let\leq=\leqslant \let\le=\leqslant
       \let\ge=\geqslant
    \fi
  \fi
\fi 

\ifnfssone
  \newmathalphabet{\mathit}
  \addtoversion{normal}{\mathit}{cmr}{m}{it}
  \addtoversion{bold}{\mathit}{cmr}{bx}{it}
  \newcommand{\rmn}[1] {\mathrm{#1}}

  \newmathalphabet{\mathbfit} 
  \addtoversion{normal}{\mathbfit}{cmr}{bx}{it}
  \addtoversion{bold}{\mathbfit}{cmr}{bx}{it}
  \newmathalphabet{\mathbfss} 
  \addtoversion{normal}{\mathbfss}{cmss}{bx}{n}
  \addtoversion{bold}{\mathbfss}{cmss}{bx}{n}
  \ifAMStwofonts
    \ifCUPmtlplainloaded \else
      \UseAMStwoboldmath
      \makeatletter
      \new@mathgroup\upmath@group
      \define@mathgroup\mv@normal\upmath@group{eur}{m}{n}
      \define@mathgroup\mv@bold\upmath@group{eur}{b}{n}
      \edef\UPM{\hexnumber\upmath@group}
      \new@mathgroup\amsa@group
      \define@mathgroup\mv@normal\amsa@group{msa}{m}{n}
      \define@mathgroup\mv@bold\amsa@group{msa}{m}{n}
      \edef\AMSa{\hexnumber\amsa@group}
      \makeatother
      \mathchardef\upi="0\UPM19
      \mathchardef\umu="0\UPM16
      \mathchardef\upartial="0\UPM40
      \mathchardef\leqslant="3\AMSa36
      \mathchardef\geqslant="3\AMSa3E

      \let\leq=\leqslant \let\le=\leqslant
       \let\ge=\geqslant
    \fi
  \fi
\fi 

\ifnfsstwo
  \newcommand{\rmn}[1] {\mathrm{#1}}

  \DeclareMathAlphabet{\mathbfit}{OT1}{cmr}{bx}{it}
  \SetMathAlphabet\mathbfit{bold}{OT1}{cmr}{bx}{it}
  \DeclareMathAlphabet{\mathbfss}{OT1}{cmss}{bx}{n}
  \SetMathAlphabet\mathbfss{bold}{OT1}{cmss}{bx}{n}
  \ifAMStwofonts
    \ifCUPmtlplainloaded \else
      \DeclareSymbolFont{UPM}{U}{eur}{m}{n}
      \SetSymbolFont{UPM}{bold}{U}{eur}{b}{n}
      \DeclareSymbolFont{AMSa}{U}{msa}{m}{n}
      \DeclareMathSymbol{\upi}{0}{UPM}{"19}
      \DeclareMathSymbol{\umu}{0}{UPM}{"16}
      \DeclareMathSymbol{\upartial}{0}{UPM}{"40}
      \DeclareMathSymbol{\leqslant}{3}{AMSa}{"36}
      \DeclareMathSymbol{\geqslant}{3}{AMSa}{"3E}

      \let\leq=\leqslant \let\le=\leqslant
       \let\ge=\geqslant
    \fi
  \fi
\fi 

\ifCUPmtlplainloaded \else
  \ifAMStwofonts \else 
    \def\upi{\pi}
    \def\umu{\mu}
    \def\upartial{\partial}
  \fi
\fi

\title{Early-type galaxies in low-density environments} 

\author[Harald Kuntschner et~al.]{Harald Kuntschner$^{1,2}$, Russell J.
  Smith$^{1,3}$, Matthew Colless$^4$, Roger L. Davies$^1$, \cr
  Raven Kaldare$^{4,5}$, Alexandre Vazdekis$^{6}$\\
~\\
  $^1$ University of Durham, Department of Physics, South Road, Durham
  DH1 3LE, UK\\
  $^2$ European Southern Observatory, Karl-Schwarzschild-Str. 2, 85748
  Garching, Germany (present address)\\
  $^3$ Departamento de Astronom\'\i a y Astrof\'\i sica, P. Univ.
  Cat\'olica de Chile, Casilla 306, Santiago 22, Chile (present
  address)\\
  $^4$ Research School of Astronomy \& Astrophysics, The Australian
  National University, Weston Creek, ACT 2611, Australia\\
  $^5$ Institute of Astronomy, University of Cambridge, Madingley Road,
  Cambridge CB3 0HA, UK \\
  $^6$ Instituto de Astrofisica de Canarias, 38200 La Laguna, Tenerife, Spain}

\date{submitted 14.12.2001; accepted 23.07.2002}

\pagerange{\pageref{firstpage}--\pageref{lastpage}}
\pubyear{2002}

\begin{document}

\maketitle

\label{firstpage}

\begin{abstract}
  We describe the construction and study of an objectively-defined
  sample of early-type galaxies in low-density environments. The sample
  galaxies are selected from a recently-completed redshift survey using
  uniform and readily-quantified isolation criteria, and are drawn from
  a sky area of $\sim$700 deg$^2$, to a depth of 7\,000~\kms\/ and an
  apparent magnitude limit of $b_J \le 16.1$. Their early-type (E/S0)
  morphologies are confirmed by subsequent CCD imaging. Five out of the
  nine sample galaxies show signs of morphological peculiarity such as
  tidal debris or blue circumnuclear rings. We confirm that E/S0
  galaxies are rare in low-density regions, accounting for only
  $\approx$8\% of the total galaxy population in such environments.  We
  present spectroscopic observations of nine galaxies in the sample,
  which are used, in conjunction with updated stellar population
  models, to investigate star-formation histories.  Our line-strength
  analysis is conducted at the relatively high spectral resolution of
  4.1~\AA.  Environmental effects on early-type galaxy evolution are
  investigated by comparison with a sample of Fornax cluster E/S0s
  (identically analysed). Results from both samples are compared with
  predictions from semi-analytic galaxy formation models. From the
  strength of \oii\ emission we infer only a low level of ongoing star
  formation ($<0.15$~M$_\odot$\,yr$^{-1}$). Relative to the Fornax
  sample, a larger fraction of the galaxies exhibit \oiiib\/ nebular
  emission and, where present, these lines are slightly stronger than
  typical for cluster E/S0s.  The Mg--$\sigma$ relation of E/S0s in
  low-density regions is shown to be indistinguishable from that of the
  Fornax sample.  Luminosity-weighted stellar ages and metallicities
  are determined by considering various combinations of line-indices;
  in particular the \HgF\/ {\em vs}\/ Fe5015 diagram cleanly resolves
  the age--metallicity degeneracy at the spectral resolution of our
  analysis. At a given luminosity, the E/S0 galaxies in low-density
  regions are younger than the E/S0s in clusters (by $\sim$2--3~Gyr),
  and also more metal-rich (by $\approx$0.2~dex). We infer that an
  anti-correlation of age and metallicity effects is responsible for
  maintaining the zero-point of the Mg--$\sigma$ relation. The youngest
  galaxies in our sample show clear morphological signs of interaction.
  The lower mean age of our sample, relative to cluster samples,
  confirms, at least qualitatively, a robust prediction of hierarchical
  galaxy formation models. By contrast, the enhanced metallicity in the
  field is contrary to the predictions and highlights shortcomings in
  the detailed treatment of star-formation processes in current models.
  The [Mg/Fe] abundance ratio appears to span a similar, mostly
  super-solar, range both in low-density regions and in Fornax cluster
  galaxies. This result is quite unexpected in simple hierarchical
  models.
\end{abstract}


\section{INTRODUCTION}
\label{sec:intro}
Hierarchical galaxy formation models predict significantly different
formation histories for early-type galaxies in cluster and low-density
environments (Baugh, Cole \& Frenk 1996, Kauffmann \& Charlot 1998). In
these models, present day clusters form from the highest peaks in the
primordial density fluctuations where major mergers of dark matter
halos, which harbour the first galaxies, rapidly produce
bulge-dominated galaxies at high redshifts ($z\ge2$).  The merging of
galaxies and the infall of new cold gas cannot continue once the
relative velocity dispersion of galaxies becomes large ($\ge
500$~\kms), \ie\/ the deep potential well of a cluster has been formed.
Within this scenario it is possible to reconcile the characteristic
ingredient of hierarchical galaxy formation, the merging process, with
the observational finding that most stars in luminous cluster
elliptical galaxies formed at $z \ge 2$ (\eg\/ Aragon-Salamanca et~al.
1993, Ellis et~al. 1997, van Dokkum et~al. 1998).

\begin{figure}
\psfig{file=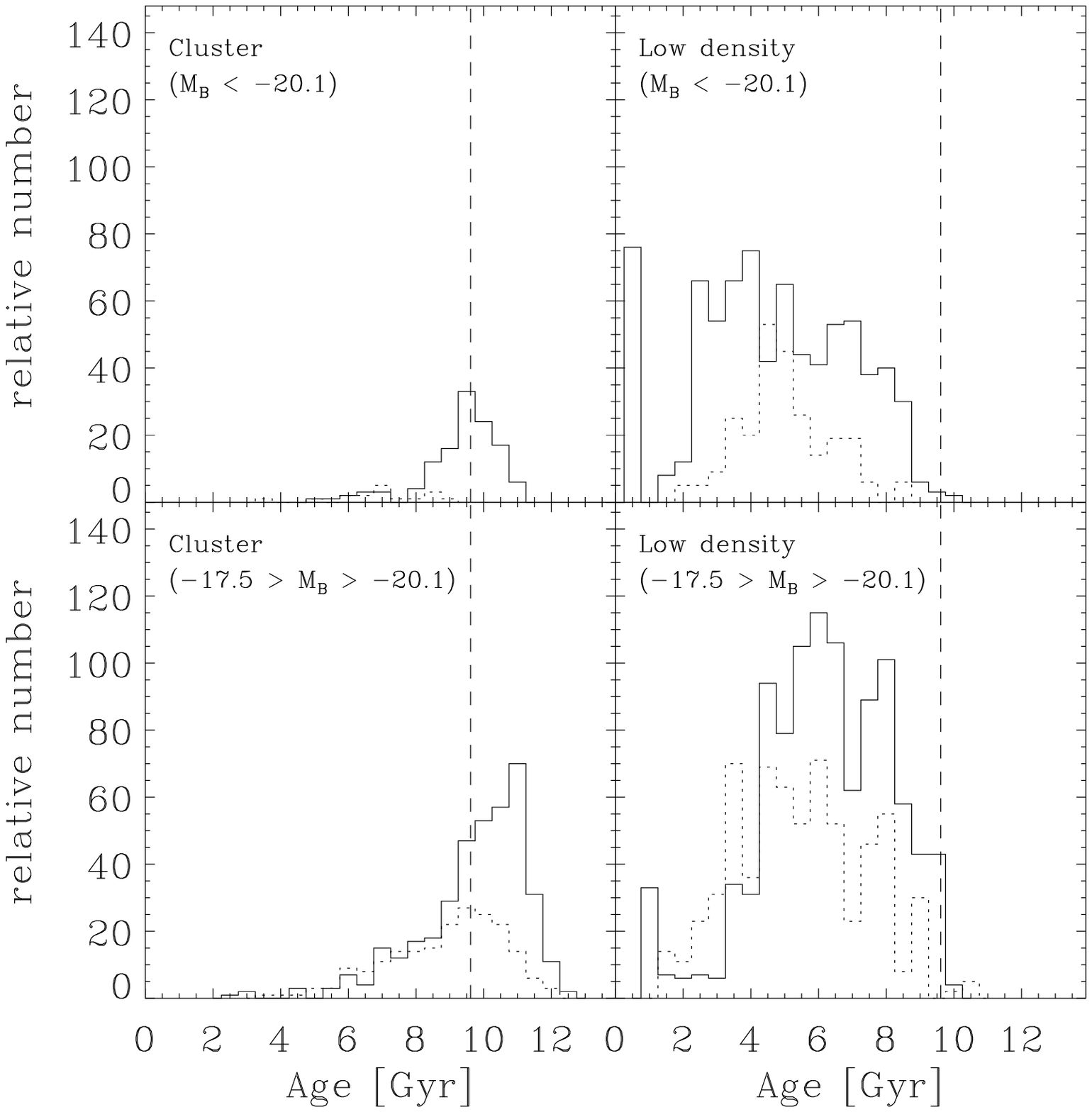,width=8.5cm}
\caption[]{\label{fig:cdm_carlton}
  Distributions of luminosity-weighted ages for early-type galaxies at
  $z=0$, predicted by hierarchical galaxy formation models
  \cite{bau96,col00}. The four panels show predictions for faint and
  bright early-type galaxies in clusters (\ie\/ residing in dark matter
  halos with $>10^{14}$~M$_\odot$) and low-density environments (\ie\/
  in dark matter halos with $<10^{13}$~M$_\odot$). Ellipticals and
  lenticular galaxies are represented by solid lines and dotted lines,
  respectively. The vertical dashed lines indicate the mean age of
  cluster ellipticals (\ie\/ 9.6~Gyr). For the predictions we adopt a
  cosmology of $H_0=72$~\kmsM\/ and $q_0=0.3$.}
\end{figure}

In low-density regions, hierarchical models predict that (i) galaxies
can accrete new cold gas and perhaps build up a stellar disk driving
the morphology towards later types and (ii) that major mergers continue
to take place at redshifts well below unity. As a result of this,
early-type galaxies in low-density regions are able to incorporate
stars formed at low redshifts and therefore should have, at the present
day, younger luminosity-weighted ages than the equivalent cluster
population \cite{bau96,kau98,gov99,col00}. As shown in
Figure~\ref{fig:cdm_carlton} the models predict for cluster ellipticals
a mean luminosity-weighted age of 9.6~Gyr (dashed line). The cluster
S0s are predicted to be $\sim$1~Gyr younger. Both ellipticals and
lenticular galaxies in clusters show a weak trend in the sense that
fainter galaxies are older. By contrast, the models predict that
early-type galaxies in low-density regions should show a broad age
distribution over the whole luminosity range with a mean age of
5--6~Gyr.

Recent observational efforts to investigate the formation of early-type
galaxies in low-density regions, and test the model predictions
discussed above, have largely focused on HST imaging of galaxies at
redshifts $0.1 \le z \le 1$. When early-type galaxies (`spheroidals')
are selected by morphology alone, no significant evidence of a decline
in comoving number density with look-back time is found \cite{men99}.
At face value this suggests an early formation epoch.  However, some of
these galaxies show colours which are too blue to be consistent with
the predictions of a simple high-redshift monolithic-collapse model
\cite{men99}.  Furthermore, there is evidence that the dispersion in
colour among field ellipticals is larger than for the equivalent
cluster population at $z \simeq 0.55$ (Schade et~al.  1999; see also
Larson, Tinsley \& Caldwell 1980 for a low redshift analogy). Schade
et~al. also found spectroscopic evidence of ongoing star formation in
the field, since about one-third of the field elliptical galaxies show
\oii\/ lines with equivalent widths in excess of 15 \AA. These results
are further supported by the study of Menanteau, Abraham \& Ellis
(2001) who investigated the {\em internal}\/ colour variations of faint
spheroidals in the HDFs. They find that at least one third of the
galaxies show strong variations in internal colour, mostly showing
centrally located blue cores, and conclude that at $z \simeq 1$
approximately 50\% of the field spheroidals experience episodes of star
formation.

Recent fundamental plane studies of field early-type galaxies
\cite{vdok01,treu01} find, within their observational errors, no
significant difference between cluster and field samples at $z \le
0.5$. However, both samples provide evidence that the stars in field
early-types are marginally younger than the equivalent cluster
population, although most of the stars must have formed at high
redshifts. van Dokkum et~al. conclude that their measurement of the
evolution of the $M/L_B$ ratio with redshift is inconsistent with the
predictions of semi-analytical models for galaxy formation
\cite{dia01}. While the models predict a systematic offset between
field and cluster in $M/L_B$ at all redshifts (the field being brighter
at a given mass), van Dokkum et~al. find no significant offset in their
data.

There are a few investigations of the nearby population of early-type
galaxies in low-density regions. de~Carvalho \& Djorgovski (1992)
investigate the properties of field and cluster early-type galaxies
using a subset of the `7~Samurai' sample \cite{fab87} and the data from
Djorgovski \& Davis (1987). They conclude that field ellipticals show
more scatter in their parameters than cluster galaxies indicating the
presence of younger stellar populations in the field.  Silva \& Bothun
(1998) investigate a sample of nearby early-type galaxies, specifically
including galaxies with disturbed morphologies such as shells and tidal
tails (see also Schweizer \& Seitzer 1992). From their analysis of
near-IR colours they conclude that there is little or no evidence for
an intermediate age (1--3~Gyr) population of significant mass ($>$10\%)
in their sample, irrespective of morphological details. Colbert,
Mulchaey \& Zabludoff (2001) have undertaken an imaging survey of 23
nearby isolated early-type galaxies, finding morphological evidence for
recent merging (\eg\/ shells and tidal features) in 41\% of the
galaxies, as compared to only 8\% in their comparison sample of group
members. Bernardi et~al. (1998) investigate the Mg$_2$--$\sigma$
relation in a large sample of early-type galaxies drawn from the ENEAR
survey \cite{dcos00} and find that there is a small difference in the
zero-point between cluster and field galaxies. They interpret this
offset as an age difference, in the sense that field galaxies are
younger by $\sim$1~Gyr. They conclude however, that the stars in both
field and cluster early-type galaxies formed mostly at high redshifts.

One of the main obstacles for studies of field galaxies is the exact
treatment of the selection process. There are many possible definitions
for the term `field', and it is critical to account for the different
selection criteria when comparing published studies. For example, while
Colbert et~al. find only 30 isolated early-type galaxies in the RC3 (de
Vaucouleurs 1991, within $cz < 9\,900$~\kms), the field sample of
Bernardi et~al. comprises more than two thirds of the entire ENEAR
catalogue (631 out of 931 galaxies within $cz < 7\,000$~\kms).
Clearly, the definitions of what is a field galaxy differ widely even
for nearby galaxy samples. It is even more difficult to compare medium-
or high-redshift samples, where redshift data is sparse, and selection
criteria for the field often ill-defined.

The present paper presents a high-quality spectroscopic study of the
stellar populations of early-type galaxies in low-density environments.
In Section~\ref{sec:sample} we describe the precise and reproducible
selection criteria according to which our new sample is selected. The
observations and basic data reduction processes are outlined in
Section~\ref{sec:data}, in which we also describe the verification and
refinement of the galaxy sample. Our measurements of absorption (and
emission) line strengths are presented in
Section~\ref{sec:linestrengths}. The principal results, detailed in
Section~\ref{sec:results}, derive from analysis of (i) the Mg--$\sigma$
relation, (ii) the luminosity-weighted ages and metallicities and (iii)
the [Mg/Fe] abundance ratios. We discuss these results in
Section~\ref{sec:discussion}, relating them to previous studies, and
comparing with the expectations from hierarchical scenarios for galaxy
formation. Our conclusions are presented in
Section~\ref{sec:conclusion}.

Throughout this paper, for model predictions as well as observations,
luminosities and physical scales are computed for a Hubble constant
$H_0=72$~\kmsM. The adopted deceleration parameter is $q_0=0.3$; using
a negative $q_0$, as preferred by SN~Ia data \cite{riess98,perl99},
would have negligible effects on these calculations.

\section{Sample Selection}
\label{sec:sample}
An important limitation of previous work has been the difficulty in
selecting a genuine sample of nearby early-type galaxies in low-density
environments. The recent FLASH (FLAIR-Shapley-Hydra) redshift survey of
Kaldare et~al. (2001\footnote{Strictly, this paper is based on a
  preliminary version of the FLASH redshift catalogue, as of December
  1998.}) provides a good basis for constructing such a sample, with a
luminosity range similar to that in typical nearby cluster samples
($M_B \le -17.5$).  The survey provides redshifts for 2931 of the 4737
galaxies brighter than $b_J$=16.7 in a strip of sky covering 10\degr\ 
in Galactic latitude ($b$=25\degr--35\degr) and 70\degr\ in Galactic
longitude ($l$=260\degr--330\degr). The survey spans the region between
the Shapley Concentration and the Hydra cluster and reaches out to
beyond 20\,000~\kms\/ (median redshift $\approx 9\,800$~\kms\/). The
source catalogue is based on the Hydra-Centaurus Catalogue of
Raychaudhury (1989, 1990) and is derived from APM scans of Southern Sky
Survey plates. The $b_J$ magnitude system is defined by the IIIa-J
emulsion of the plates and the GG395 filter, and is related to the
standard Johnson B and V magnitudes by $b_J$=B-0.28(B$-$V) for
$-$0.1$\leq$(B$-$V)$\leq$1.6 (Blair \& Gilmore 1982). The redshifts in
the survey catalogue comprise both new measurements obtained with the
FLAIR fibre spectrograph on the United Kingdom Schmidt Telescope, of
the Anglo-Australian Observatory, and measurements from the literature;
they have an rms precision of approximately 60~\kms. The completeness
of the survey for early-type galaxies is a function of magnitude $m$
and is approximately given by $f(m)=1$ for $m\leq13.5$ and $f(m) =
-5.48 + 1.00m - 0.039m^2$ for $m>13.5$. Therefore, at our limiting
magnitude of $b_J = 16.1$ (equivalent to $B \approx 16.3$ for
early-type galaxies) the completeness is still 51\%.

For the selection of our low-density region sample of early-type
galaxies we chose a redshift limit of $cz \le 7\,000$~\kms, an apparent
magnitude limit of $b_J \le 16.1$ (no limit on the brightest galaxies),
and required $\le2$ neighbours ($b_J \le 16.7$) within a
redshift-scaled search radius corresponding to 0.8\degr\/ at $cz =
7\,000$~\kms\/ (\ie\/ a radius of 1.3 Mpc in our adopted cosmology) and
a constant depth of $\pm 350$~\kms. Within $cz \le 7\,000$~\kms\/ the
survey contains 1069 galaxies. Of these, 237 galaxies are classified as
early-types and brighter than $b_J=16.1$ (after excluding those
galaxies which are so close to edges of the survey that the search
radius is not fully sampled).

These selection criteria yielded a sample of $40$ E and S0 galaxies in
low-density regions. Visual inspection of `Digitized Sky Survey' (DSS)
images, however, revealed that some of the galaxies have later
morphology than indicated by Raychaudhury's (1989) classification.
Excluding these galaxies resulted in a sample of 30 galaxies of which
24 were observed (see Section~\ref{sec:data}). We note that a visual
inspection (DSS images) of galaxies satisfying our isolation criteria
and classified as spiral by Raychaudhury yielded only four additional
galaxy targets which can be classified as S0. Non of these galaxies
were observed. Since not all of the galaxies are included in well-known
catalogues we refer to them in this paper as LDR\,xx where LDR stands
for Low Density Region (see Table~\ref{tab:fieldsample} and
\ref{tab:nonldrsample}).

\section{THE OBSERVATIONS AND BASIC DATA REDUCTION}
\label{sec:data}
\subsection{Observational techniques}
Spectroscopic observations were obtained during two runs (1999 January
18--20 and 1999 February 11--14) at the ANU 2.3m telescope at Siding
Spring Observatory, Australia. The first run yielded only one night of
usable data; the second run was entirely clear. Both runs used
identical instrumentation. The observations were made with the Double
Beam Spectrograph \cite{rod88}, in which a dichroic feeds red and blue
beams to two identical spectrographs. The analyses presented in this
paper, however, are based on the blue-beam spectra, and so the details
below refer to the blue-beam instrumentation only. A
600~line\,mm$^{-1}$ grating yielded a spectral resolution of
$\sim$2.6~\AA\/ (FWHM, instrumental velocity resolution $\sigma_{\rm
  instr}\simeq 65$~\kms\/ at 5200~\AA), and a sampling of
$1.1$~\AA\/\,pixel$^{-1}$. The spatial scale was
0.91~arcsec\,pixel$^{-1}$ and the slit width was 2~arcsec (aligned
roughly along the major axis of each galaxy). The detector was a SITe
CCD of 1752$\times$532 pixel format. The resulting spectra cover the
rest-wavelength range 3690--5600~\AA, including many prominent Balmer
and metallic lines (\Hd, \Hg, \Hb, \mgb, Fe5270, Fe5335 amongst others)
and the wavelengths of nebular \oiia\/ and \oiii\/ emission.
Table~\ref{tab:instr_msso} summarizes the instrumental set-up.

\begin{table}
  \caption[]{The instrumental set-up.}
  \label{tab:instr_msso}
  \begin{tabular}{ll} \hline
    Telescope            &MSSSO (2.3m)                          \\
    Dates                &18-20 Jan 1999, 11-14 Feb 1999        \\
    Instrument           &DBS spectrograph (blue beam)          \\ \hline
    Spectral range       &3690-5600 \AA                         \\
    Grating              &600 line\,mm$^{-1}$                   \\
    Dispersion           &1.1~\AA\,pixel$^{-1}$                 \\
    Resolution (FWHM)    &$\sim$2.6~\AA                         \\
    Spatial Scale        &0\farcs91~pixel$^{-1}$                \\
    Slit Width           &2\farcs0                              \\
    Detector             &SITe ($1752 \times 532$ pixels; $15 \times 15\, \umu m$) \\
    Gain                 &$1.0\ e^- \mbox{ADU}^{-1}$            \\
    Read-out-noise       &$5.0\ e^-$ (rms)                      \\
    Typical seeing       &$\sim$2\arcsec                        \\ \hline
  \end{tabular}
\end{table}

The standard calibration frames were obtained, including zero-exposure
bias frames, tungsten lamp exposures for flat-fielding and twilight-sky
exposures for vignetting corrections. Neon--argon arc lamp spectra were
obtained to provide wavelength calibration. Arc exposures were made
immediately prior to, or subsequent to, each exposure. Long
integrations were divided into exposures of at most 1200~s, each with a
separate arc observation, so as to track any flexure in the
telescope--spectrograph system during the exposure.

In order to calibrate the two runs and provide templates for redshift
and velocity dispersion measurements we observed stars selected from
the Lick stellar library catalogue \cite{wor94b}, covering a broad
range of spectral types. Spectrophotometric standard stars were also
observed to calibrate the response function of the system.

For the target galaxies, the total exposure times were typically
2400~s or 3600~s. In two cases (LDR\,15 \& LDR\,27) the surface
brightness of the galaxy was too low to obtain spectra of sufficient
signal-to-noise ratio in a reasonable integration time (\eg\/
$S/N\ga40$ per \AA\ in $\la$1.5~hr). For these galaxies, we obtained a
single exposure, in order to make a rough spectral classification from
the presence and strength of emission lines.

\begin{figure}
\psfig{file=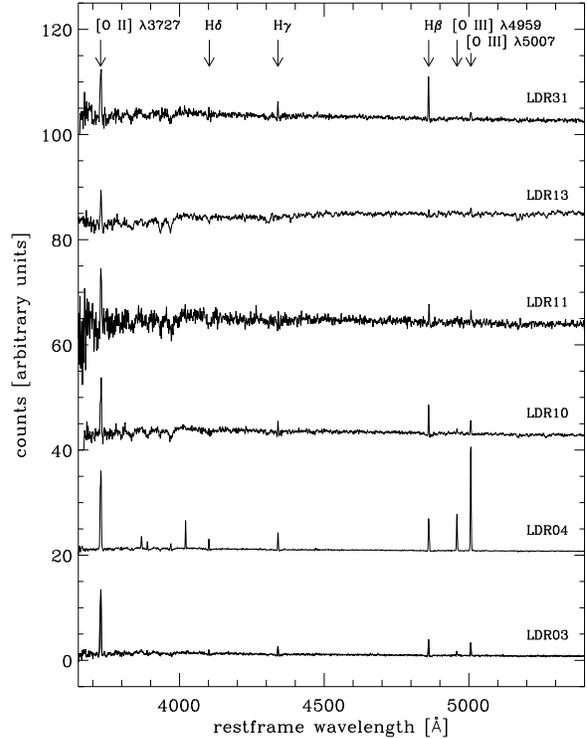,width=8.5cm}
\caption[]{\label{fig:spiral_spec}Spectra of galaxies which
  were rejected from the final sample.  These galaxies all exhibit
  spiral morphologies, generally noticed on the auto-guide camera during
  the spectroscopic observations. LDR\,13 was rejected after CCD
  imaging was obtained (for details see text). The spectra are shown at
  rest-wavelength. Important emission features are indicated at the top
  and the galaxy identification is given to the right hand side of each
  spectrum.}
\end{figure}

For five galaxies we noticed a spiral morphology on the CCD auto-guider
and took only one 20~min exposure. These galaxies (LDR\,03, LDR\,04,
LDR\,10, LDR\,11 \& LDR\,31) show emission line spectra typical for
spiral galaxies (see Figure~\ref{fig:spiral_spec}) and were excluded
from our final sample.

For calibration purposes, five comparison galaxies were observed in
overlap with the sample of Kuntschner (2000): NGC\,1381, NGC\,1399,
NGC\,1404 and IC\,2006 (all in the Fornax cluster), plus the Leo-group
member NGC\,3379. We also obtained spectra of three early-type galaxies
in the Fornax cluster which were not previously observed by Kuntschner
(NGC\,1344, NGC\,1366 and NGC\,1387). For a summary of the comparison
galaxies in the Fornax cluster, see Table~\ref{tab:compsample}.

\subsection{Basic data reduction}
Data reduction followed standard methods, and was performed using a
combination of the {\sc starlink} and {\sc iraf} software packages.
Frames were de-biased (allowing for a small along-the-slit bias-level
variation, which changes slightly from exposure to exposure),
flatfielded to remove pixel-to-pixel sensitivity variations and
corrected for vignetting along the slit. Cosmic ray events were removed
from the frames interactively, and with great care, using the {\tt
  lineclean} task within {\sc iraf}. Wavelength calibration was
achieved by fitting a low-order (usually cubic) polynomial to $\sim$40
lines in the arc spectra. The typical rms scatter in the calibration
was 0.07~\AA.

From each exposure, one-dimensional spectra were extracted for a
central aperture. The radial gradients of line-strengths are not known
a priori for individual galaxies. It is therefore desirable to scale
the extraction aperture with redshift, so as to probe a roughly
constant physical scale at all of the galaxies in the sample.
J{\o}rgensen, Franx \& Kj{\ae}rgaard (1995, see also
Appendix~\ref{sec:aperture}) have given a recipe for converting
rectangular aperture dimensions to an equivalent circular aperture
diameter. We scale the along-the-slit extraction length with redshift
such that this equivalent circular aperture has 1.08~kpc diameter for
all galaxies. The extraction area is $4 \arcsec \times 2 \arcsec$ at
the characteristic depth of the sample (5\,000~\kms).

Finally, the individual spectra for each target were added to yield the
final spectrum from which redshift, velocity dispersion and
line-indices are measured. Prior to the index measurements the
continuum shape was corrected to a relative flux scale using the
spectrophotometric standard GD\,108 \cite{oke90}.

\subsection{Redshift determination}
\label{sec:redshifts}
In a first analysis, redshifts were measured for the galaxies in
low-density regions with the {\sc IRAF} routine {\tt fxcor} and corrected to
heliocentric values (see Table~\ref{tab:fieldsample}). A comparison of
these redshifts with the catalogue of Kaldare et~al. (2001) showed that
our high-quality measurements disagree with the catalogue by more than
three times the expected 1$\sigma$ error for eight galaxies. In order
to test whether these galaxies still satisfy our isolation criteria we
re-ran our selection program using the new redshifts.

This resulted in three galaxies being removed, having redshifts beyond
7\,000~\kms\/ (our redshift limit). Four galaxies were reclassified as
group/cluster members (see Table~\ref{tab:nonldrsample}). We note that
by searching a given redshift survey for isolated early-type galaxies,
one is likely to pick up many of the erroneous redshifts in the survey,
since a large error in $z$ most likely moves these galaxies from
clusters and groups (where they are most common) to lower-density
environments (which occupy most of the survey volume).

\subsection{Imaging}
\label{sec:imaging}
We have obtained optical ($UVR$) and near-infrared ($K_s$) imaging data
at the CTIO 1.5m telescope. The morphologies of our sample galaxies,
estimated from visual inspection of the optical images, are best
described as E or S0. Many of the galaxies, however, present
peculiarities (tidal tails or debris, rings, secondary
intensity-maxima, disturbed companions, etc.) which are likely related
to merger and/or interaction events (indeed six of the nine galaxies in
the final sample are listed in the Arp \& Madore (1987) catalogue of
peculiar galaxies).

The sample galaxy LDR\,13 is an exception to the above statement. It
presents an exponential profile and, after subtraction of an
elliptical-isophote model, spiral arms are clearly visible. This galaxy
was thus reclassified as a spiral galaxy, and removed from later
consideration. A spectrum of LDR\,13 is shown in
Figure~\ref{fig:spiral_spec}.

Our broad classifications based on visual inspection are included in
Table~\ref{tab:fieldsample}. A more detailed and quantitative
description of the morphological characteristics of the individual
sample galaxies, and their relation to the spectroscopic results, will
be provided in a forthcoming paper.

\begin{table*}
  \caption{Catalogue of early-type galaxies in low-density regions}
  \label{tab:fieldsample}
  \begin{tabular}{lllcccccrrc} \hline
Name  & ESO cat.        & AM cat.     & Type        & \multicolumn{2}{c}{RA~~J2000~~DEC}                 & $b_J$& $M_B$   & $\sigma$     & $cz_{helio}$  & neigh. \\
      &                 &             &             &                        &                           & [mag]& [mag]   & [\kms]       & [\kms]        &\\
 (1)  &  (2)            &    (3)      & (4)         &  (5)                   &  (6)                      & (7)  &   (8)   &   (9)        &     (10)      &  (11) \\ \hline
LDR\,08 & ESO\,503-G005 &AM\,1112-272 & S0   (S...) &  11$^h$15$^m$15\farcs6 &-27\degr39$^m$38\arcsec    &15.30 &  -18.5  &$ 82.0\pm 4.4$&   3\,880$\pm$10 & 0/1 \\
LDR\,09 & ESO\,503-G012 &AM\,1115-255 & E    (S0)   &  11$^h$17$^m$50\farcs9 &-26\degr08$^m$04\arcsec    &14.76 &  -17.6  &$146.8\pm 2.6$&   2\,138$\pm$10 & 2/2 \\
LDR\,14 & ESO\,379-G026 &AM\,1203-354 & E    (S0)   &  12$^h$06$^m$16\farcs9 &-35\degr58$^m$51\arcsec    &14.24 &  -19.5  &$139.8\pm 2.4$&   3\,903$\pm$10 & 1/1 \\
LDR\,19 & ESO\,442-G006 &-            & S0   (S0)   &  12$^h$34$^m$06\farcs2 &-31\degr13$^m$00\arcsec    &15.42 &  -19.2  &$173.6\pm 3.8$&   5\,781$\pm$18 & 0/0 \\
LDR\,20 & ESO\,381-G004 &-            & S0   (Sa)   &  12$^h$39$^m$09\farcs6 &-34\degr46$^m$51\arcsec    &15.07 &  -18.9  &$181.9\pm 5.3$&   4\,263$\pm$14 & 1/1 \\
LDR\,22 & ESO\,382-G016 &AM\,1310-362 & E    (S0)   &  13$^h$13$^m$12\farcs3 &-36\degr43$^m$22\arcsec    &13.54 &  -19.8  &$237.2\pm 5.7$&   3\,297$\pm$14 & 2/4 \\
LDR\,29 & ESO\,445-G056 &-            & S0   (S0)   &  13$^h$50$^m$53\farcs6 &-30\degr17$^m$20\arcsec    &14.89 &  -19.6  &$154.4\pm 2.6$&   5\,682$\pm$10 & 2/3 \\
LDR\,33 & -             &AM\,1402-285 & S0   (- )   &  14$^h$05$^m$22\farcs4 &-29\degr08$^m$27\arcsec    &15.19 &  -19.5  &$159.1\pm 3.1$&   5\,999$\pm$11 & 0/0 \\
LDR\,34 & ESO\,446-G049 &-            & S0   (S0)   &  14$^h$20$^m$14\farcs4 &-29\degr44$^m$50\arcsec    &13.71 &  -20.1  &$143.7\pm 2.0$&   3\,836$\pm$10 & 0/0 \\ \hline
  \end{tabular}

\medskip

\begin{minipage}{17.5cm}
  {\em Notes:}\/ The first column gives the name of the galaxy as used
  in this paper. If the galaxy is found in the ESO/Uppsala survey of
  galaxies \cite{lau82} then we list its ESO identification in column
  two. For those galaxies tabulated in ``A catalogue of southern
  peculiar galaxies and associations'' \cite{arp87}, the corresponding
  identification is given in column three. The fourth column denotes
  our own morphological classification followed by the ESO
  classification in brackets. The tabulated coordinates (columns five
  and six) are determined from the `Digitized Sky Survey', and should
  be accurate to $\sim$1~arcsec. Column seven shows the (total) $b_J$
  magnitude, derived from Hydra--Centaurus Catalogue (Raychaudhury
  1989). The absolute magnitude $M_B$ (column eight) was calculated
  using the redshift of each galaxy ($H_0=72$~\kmsM, $q_0=0.3$), and is
  converted from $b_J$ to $B$ using the transformation given in the
  text, assuming $B-V=0.8$. The quoted $M_B$ is corrected for galactic
  extinction using the Schlegel et~al.  (1998) maps. Column nine gives
  the central velocity dispersion $\sigma$, as measured within an
  aperture equivalent to 1.08~kpc diameter. Column ten lists our
  heliocentric redshift (measured from spectra of much higher quality
  than the FLASH survey data), while column eleven shows the number of
  neighbours detected in the FLASH catalogue by applying our isolation
  criteria (see Section~\ref{sec:sample} for details). In this final
  column, the first number refers to the results obtained using the Dec
  1998 version of the FLASH catalogue; the second number is the value
  obtained using the updated catalogue as published by Kaldare et~al.
  (2001).
\end{minipage}
\end{table*}

\subsection{The final sample}
\label{sec:final_sample}
Taking into account all of the selection criteria discussed above (see
Table~\ref{tab:selectsum} and Table~\ref{tab:selectpro}), the final
sample comprises nine early-type galaxies (3 Es, 6 S0s) in low-density
regions (see Table~\ref{tab:fieldsample}). While the sample size is
small, we emphasize that the multiple, well-defined selection criteria
guarantee that these galaxies are of E/S0 morphology, and reside in
large-scale environments of very low density.

\begin{figure*}
\epsfig{file=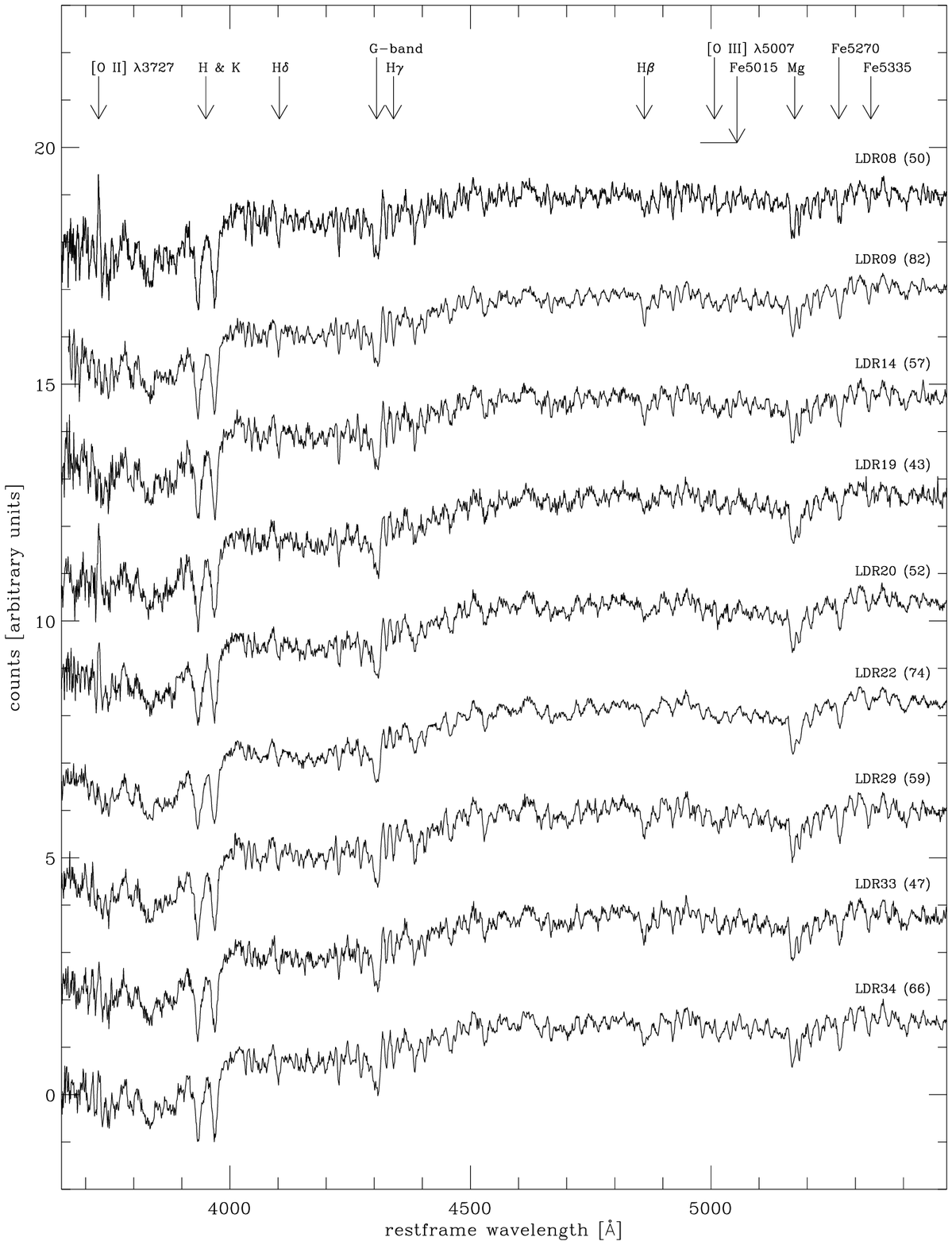,width=17.5cm}
\caption[]{\label{fig:fe_spec}Spectra of our sample of early-type
  galaxies in low-density regions. The spectra are shown at
  rest-wavelength. Characteristic emission and absorption features are
  indicated at the top of the panel and the galaxy identification is
  given to the right hand side of each spectrum. The S/N per \AA\/ is
  given in brackets after the galaxy name.}
\end{figure*}

Figure~\ref{fig:fe_spec} presents the rest-frame spectra of galaxies in
the confirmed LDR sample.  All spectra show the features characteristic
of early-type galaxies: prominent H and K lines, the G-band and the Mg
feature at $\sim$5175~\AA.  Compared to the spiral galaxy spectra in
Figure~\ref{fig:spiral_spec}, there is little emission visible in the
oxygen lines. However, about half of our sample galaxies do exhibit
weak \oii\/ emission (equivalent width between 3 and 7~\AA; detection
limit $\approx$0.5~\AA). \oiii\ emission (at 4959~\AA\ and 5007~\AA) is
rarely visible in the raw spectra, but can be distinguished after
subtraction of a model continuum. Emission line measurements are
further discussed in Section~\ref{sec:linestrengths}.

\begin{figure}
\epsfig{file=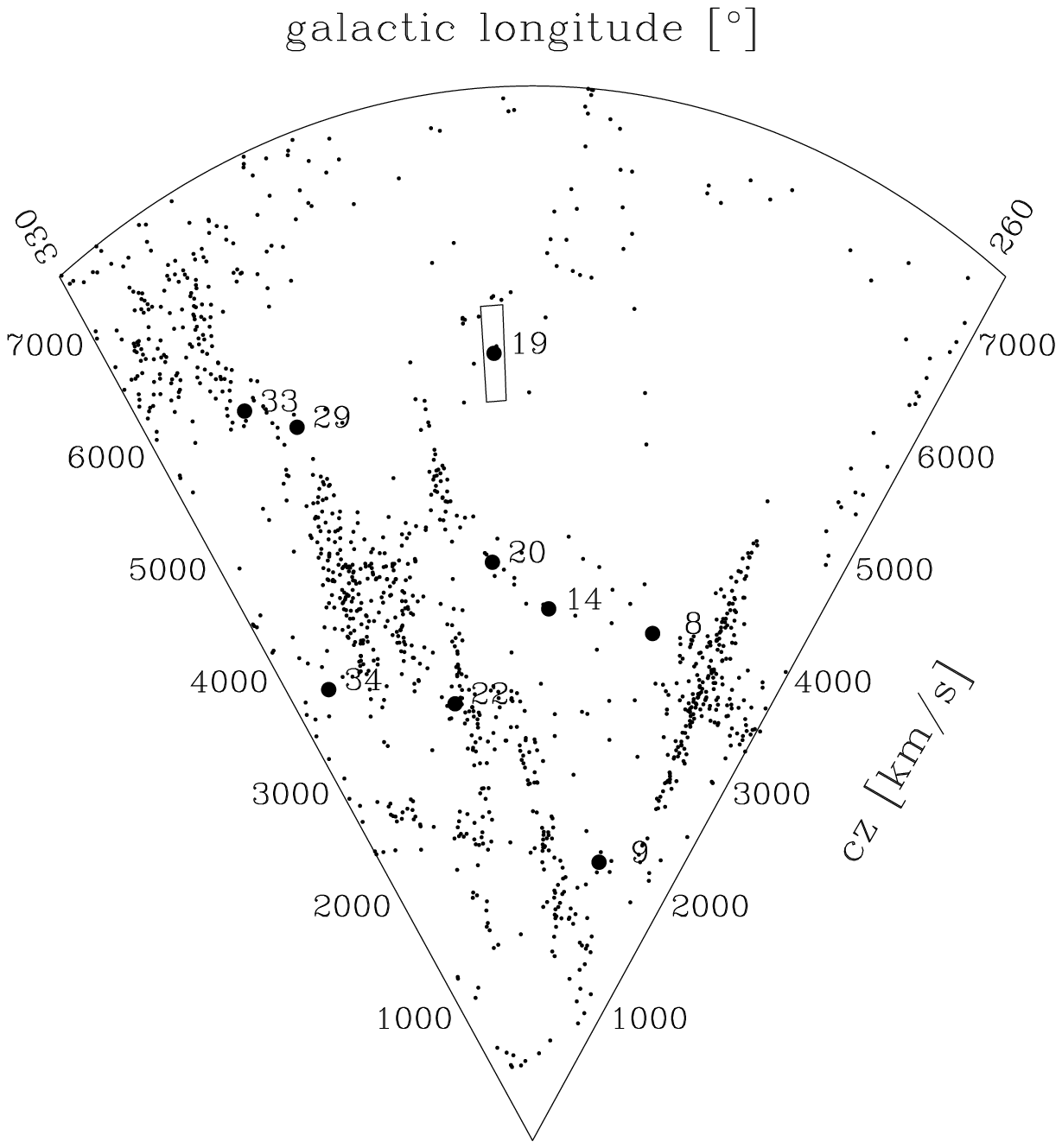,width=8.5cm,clip=,bbllx=82,bblly=19,bburx=471,bbury=436}
\caption[]{\label{fig:pos1}Redshift-space distribution of the
  FLASH-survey (Kaldare et~al. 2001) within 7\,700~\kms\/ (small dots).
  The sample of galaxies in low-density regions are indicated by filled
  circles and labelled with their catalogue numbers (see
  Table~\ref{tab:fieldsample}). The box around LDR\,19 demonstrates the
  redshift-scaled search box for neighbours around this galaxy.}
\end{figure}

\begin{figure*}
\psfig{file=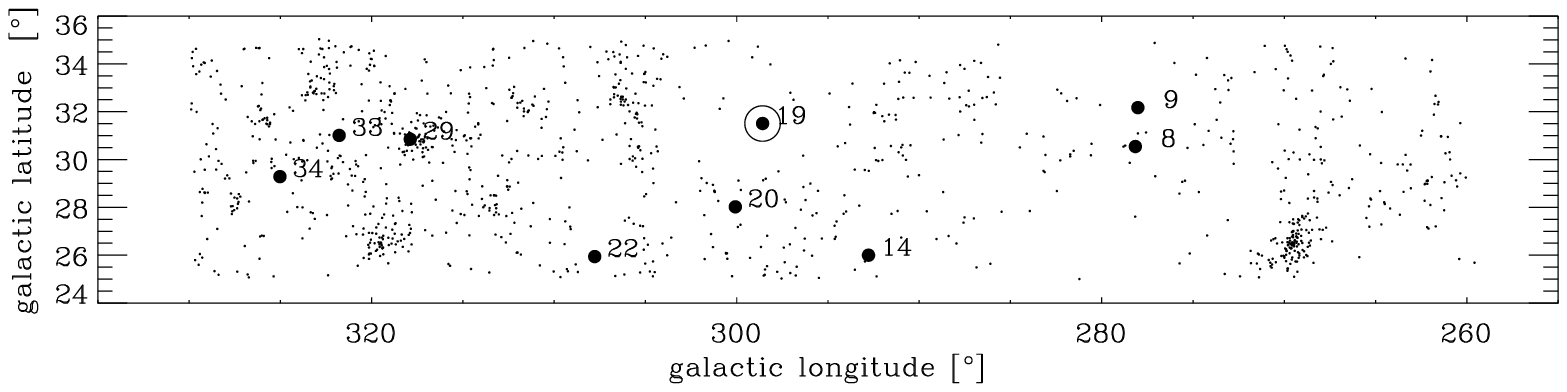,width=17.0cm}
\caption[]{\label{fig:pos2}Projected `on-sky' distribution of the
  FLASH-survey (Kaldare et~al. 2001) within 7\,700~\kms\/ (small dots).
  The sample of galaxies in low-density regions are indicated by filled
  circles. The data points are labelled with their LDR catalogue
  numbers. The circle around LDR\,19 demonstrates the redshift-scaled
  search radius for neighbours around this galaxy.}
\end{figure*}

The distribution of our sample in redshift space (`pie-diagram') and
projected on the sky are presented in Figures~\ref{fig:pos1} and
\ref{fig:pos2}, respectively. These diagrams show graphically that our
selection procedure finds galaxies well separated from dense cluster
environments such as Abell\,1060 (l,b)=(270\degr, 27\degr), Abell\,3581
(l,b)=(325\degr, 33\degr), Abell\,3574 (l,b)=(320\degr, 30\degr) and
Abell\,S0753 (l,b)=(320\degr, 26\degr) which are located within the
FLAIR redshift survey. The only sample member which might potentially
be a cluster member (albeit with extremely large relative velocity) is
LDR\,29, located $\sim$2\,000~\kms\/ beyond the mean redshift of
Abell\,3574.

\begin{figure}
\psfig{file=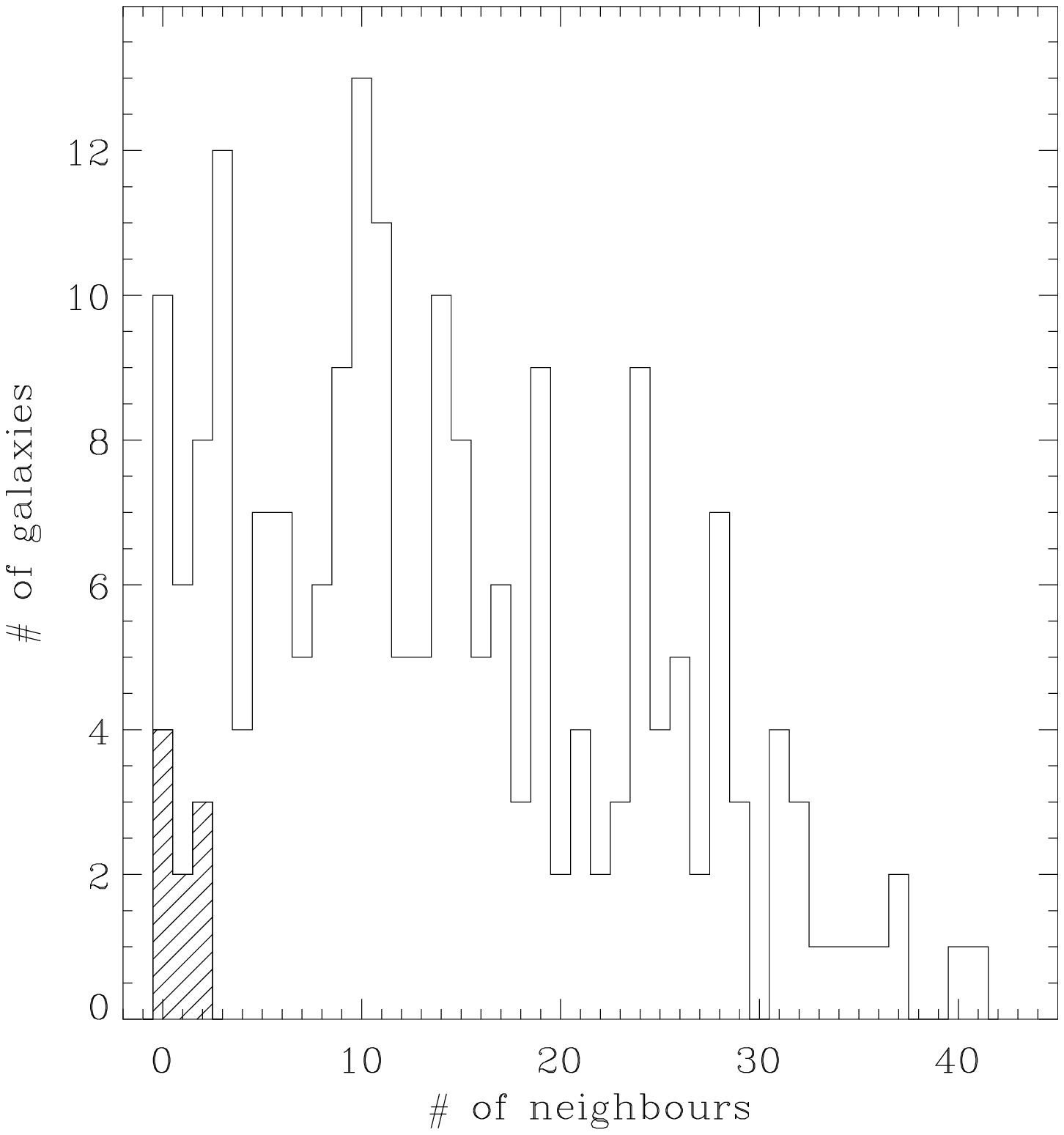,width=8.5cm}
\caption[]{\label{fig:neigbours}The distribution of `neighbour counts'
  for luminous ($M_B \le -17.5$) early-type galaxies in the FLASH
  survey within 7\,000~\kms. The number of neighbours reflects all
  galaxies (regardless of morphology and luminosity) counted within the
  redshift-scaled search box. The bin size is one. The hatched region
  indicates the location of our final sample of nine galaxies (see
  Section~\ref{sec:final_sample}).}
\end{figure}

We note that at the time of submission of this paper, an updated
version of the FLASH catalogue became available.  This catalogue
contains more literature redshifts than the version we used to select
our sample (Dec 1998). As a result LDR\,22 and LDR\,29 have 4 and 3
neighbours, respectively, in the new catalogue. This moves them
slightly out of our original selection criteria; nevertheless, the
environments of these galaxies are still best-described as being of low
density. Figure~\ref{fig:neigbours} shows the distribution of
`neighbour counts' for early-type galaxies in the FLASH survey (version
Dec 1998) within 7\,000~\kms\/ restricted to galaxies with $M_B \le
-17.5$. The hatched region indicates the location of our final sample
of nine galaxies.  Note that our experience in compiling the LDR sample
suggests that the number of isolated E/S0s is systematically
overestimated in the FLASH catalogue, due to morphological
classification and redshift errors (see Section~\ref{sec:redshifts}).
In particular, only $\sim$40\% of the observed candidates were
confirmed as fulfilling our criteria. If this effect were corrected
for, it is likely that our sample would be found to occupy the tail of
a distribution which peaks at 10-15 neighbours.

\begin{figure}
\epsfig{file=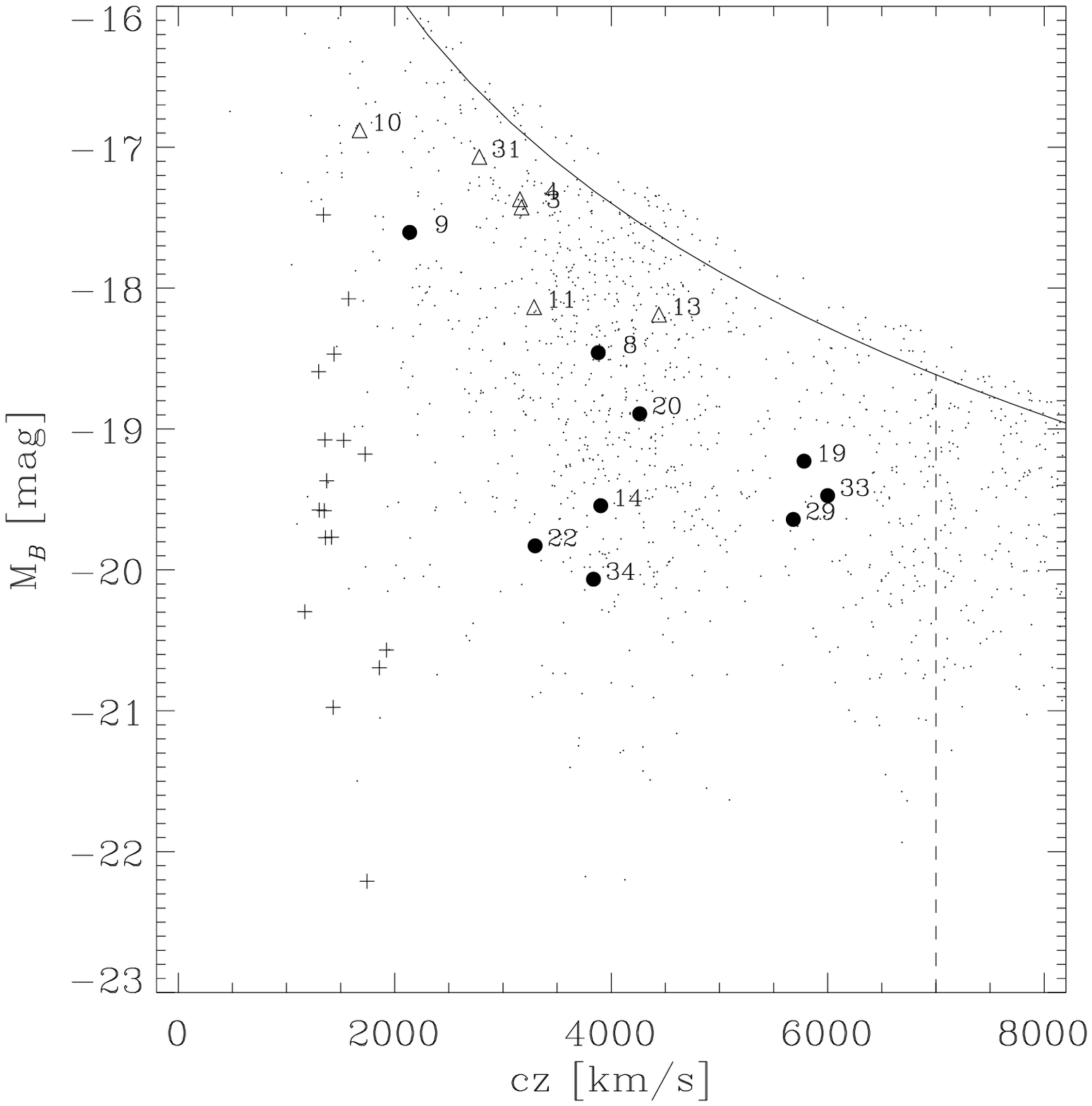,height=8.5cm,width=8.5cm}
\caption[]{\label{fig:selection} The selection function of the
  early-type galaxies in low-density regions. The small dots represent
  all galaxies in the FLASH redshift survey (Kaldare et~al. 2001).
  Filled circles represent our sample of galaxies in low-density
  regions. The open triangles show the spiral galaxies in low-density
  regions where we have spectra. The LDR catalogue numbers are
  indicated in the plot. The plus signs represent the Fornax cluster
  comparison sample. The solid line shows our faint magnitude limit of
  $b_J = 16.1$ and the dashed line indicates the redshift cutoff at
  7\,000~\kms.}
\end{figure}

Figure~\ref{fig:selection} shows the FLASH catalogue selection function
($M_B$ versus redshift diagram) with the galaxies of
Table~\ref{tab:fieldsample} highlighted. Our sample spans a luminosity
range of $M_B = -17.6$ to $-20.1$. The luminosity distribution of our
sample is similar to that of the early-type galaxies in Fornax fainter
than $M_B \simeq -20$. However, there are no galaxies in our field
sample more luminous than this (compared to 5 in Fornax,
see Section~\ref{sec:disc_selection} for further discussion).

\begin{table*}
  \caption{Observed galaxies which do not satisfy our selection criteria}
  \label{tab:nonldrsample}
  \begin{tabular}{llccccr}\hline
Name  & Name other      &  \multicolumn{2}{c}{RA~~J2000~~DEC}                & $b_J$ & $M_B$ & $cz_{helio} \pm err$ \\
      &                 &                        &                           & [mag]& [mag]   &  [\kms]       \\\hline
      &                 &                        &                           &      &         &               \\
      & \multicolumn{6}{l}{{\em Emission line and spiral galaxies in low-density regions}}\\
      &                 &                        &                           &      &         &               \\
LDR\,03 & ESO\,569-IG?002 &  10$^h$44$^m$59\farcs9 &-22\degr09$^m$09\arcsec    &15.80 &  -17.4  &   3\,171$\pm$21 \\
LDR\,04 & ESO\,501-G096   &  10$^h$46$^m$47\farcs5 &-23\degr19$^m$39\arcsec    &15.90 &  -17.4  &   3\,156$\pm$35 \\
LDR\,10 & ESO\,439-G016   &  11$^h$31$^m$51\farcs2 &-30\degr24$^m$40\arcsec    &15.02 &  -16.9  &   1\,675$\pm$12 \\
LDR\,11 & ESO\,503-G025   &  11$^h$34$^m$29\farcs0 &-26\degr52$^m$10\arcsec    &15.20 &  -18.1  &   3\,287$\pm$23 \\
LDR\,13 & -               &  12$^h$00$^m$34\farcs6 &-32\degr21$^m$50\arcsec    &15.98 &  -18.2  &   4\,441$\pm$11 \\
LDR\,31 & S753 [56]$^a$   &  14$^h$03$^m$14\farcs6 &-34\degr01$^m$17\arcsec    &15.96 &  -17.1  &   2\,781$\pm$15 \\
      &                 &                        &                           &      &         &               \\
      & \multicolumn{6}{l}{{\em Galaxies in group/cluster environments or} $cz > 7\,000$~\kms}\\
      &                 &                        &                           &      &         &               \\
LDR\,06 & ESO\,502-G005   &  10$^h$57$^m$37\farcs0 &-25\degr25$^m$39\arcsec    &14.87 &  -18.9  &   3\,901$\pm$12 \\
LDR\,17 & ESO\,441-G025   &  12$^h$16$^m$13\farcs2 &-30\degr07$^m$41\arcsec    &15.61 &  -21.0  &  14\,566$\pm$30 \\
LDR\,18 & -               &  12$^h$18$^m$40\farcs5 &-28\degr27$^m$58\arcsec    &15.87 &  -20.0  &  10\,236$\pm$12 \\
LDR\,21 & ESO\,381-G019   &  12$^h$44$^m$44\farcs1 &-35\degr52$^m$21\arcsec    &15.50 &  -20.1  &   9\,108$\pm$14 \\
LDR\,25 & ESO\,444-G038   &  13$^h$27$^m$00\farcs6 &-29\degr11$^m$30\arcsec    &14.79 &  -19.0  &   4\,153$\pm$10 \\
LDR\,26 & A\,3559 [23]$^b$&  13$^h$28$^m$05\farcs7 &-29\degr25$^m$29\arcsec    &15.75 &  -18.1  &   4\,087$\pm$12 \\
LDR\,30 & S753 [49]$^a$   &  14$^h$03$^m$07\farcs9 &-34\degr01$^m$58\arcsec    &15.70 &  -18.7  &   5\,124$\pm$22 \\
      &                 &                        &                           &      &         &               \\
      & \multicolumn{6}{l}{{\em low S/N spectra without classification}}\\
      &                 &                        &                           &      &         &               \\
LDR\,15 & -               &  12$^h$08$^m$34\farcs2 &-30\degr08$^m$53\arcsec    &16.04 &  -16.4  &   2\,157$\pm$24 \\
LDR\,27 & GSC 7269 01680  &  13$^h$31$^m$36\farcs2 &-32\degr58$^m$52\arcsec    &15.92 &  -17.6  &   3\,556$\pm$10 \\ \hline
  \end{tabular}
\medskip

\begin{minipage}{11.9cm}
  {\em Notes:}\/ $(^a)$ The cluster S753 is listed in the supplementary
  catalogue of Abell, Corwin \& Olowin (1989). The galaxy number in
  square brackets is that assigned by Willmer et~al. (1991).  $(^b)$
  The cluster Abell\,3559 is listed in Abell, Corwin \& Olowin (1989).
  The galaxy number in square brackets is that of Katgert et~al.
  (1998). The first column lists the name of the galaxy as referred to
  in this paper, column two gives the identification in the ESO/Uppsala
  survey \cite{lau82}, or other source as described above. Other
  columns are as in Table~\ref{tab:fieldsample}.
\end{minipage}
\end{table*}

\begin{table}
  \caption{Summary of our selection criteria. }
  \label{tab:selectsum}
  \begin{tabular}{ll}  \hline
    Sky area       & $l$=260\degr -- 330\degr, $b$=25\degr -- 35\degr \\
    Redshift range & $cz<7\,000$~\kms \\
    Magnitude      & $b_J\le16.1$  \\
    Morphology     & E or S0 \\
    Environment    & $\le2$ neighbours with $b_J\le16.7$ \\
                   & \ \ \ \ \ (within 1.3~Mpc projected radius and \\
                   & \ \ \ \ \  $\pm$350~\kms\/ $cz$) \\
    Spectral type  & absorption-dominated \\
    Spectral $S/N$ & $>40$ per \AA \\ \hline
  \end{tabular}
\end{table}

\begin{table}
  \caption{Sample of comparison galaxies in the Fornax cluster}
  \label{tab:compsample}
  \begin{tabular}{llcl} \hline
Name      &Type&  $\sigma \pm \sigma_{err}$ &Reference  \\ \hline
NGC\,1316 & S0 &   $224.2\pm\z2.9$ & K00 \\
NGC\,1336 & E  &  $\z75.5\pm\z5.7$ & K00 \\
NGC\,1339 & E  &   $143.7\pm\z3.7$ & K00 \\
NGC\,1344 & S0 &   $165.7\pm\z4.0$ & this study \\
NGC\,1351 & E  &   $147.9\pm\z3.5$ & K00 \\
NGC\,1366 & S0 &   $120.5\pm\z2.4$ & this study \\
NGC\,1373 & E  &  $\z77.7\pm\z9.4$ & K00 \\
NGC\,1374 & E  &   $168.8\pm\z7.7$ & K00 \\
NGC\,1379 & E  &   $116.8\pm\z3.6$ & K00 \\
NGC\,1380 & S0 &   $200.3\pm\z6.1$ & K00 \\
NGC\,1381 & S0 &   $142.8\pm\z4.2$ & K00, this study \\
NGC\,1387 & S0 &   $194.6\pm\z5.0$ & this study \\
NGC\,1399 & E  &   $348.3\pm10.9$ & K00, this study \\
NGC\,1404 & E  &   $244.9\pm\z8.2$ & K00, this study \\
NGC\,1419 & E  &   $113.7\pm\z4.1$ & K00 \\
NGC\,1427 & E  &   $170.7\pm\z2.2$ & K00 \\
IC\,2006  & S0 &   $127.9\pm\z2.6$ & K00, this study \\ \hline
  \end{tabular}

\medskip

\begin{minipage}{8cm}
  {\em Notes:}\/ K00: Line-strength and velocity dispersions previously
  published by Kuntschner (2000); All velocity dispersions are
  corrected to nominal aperture of $4 \arcsec \times 2 \arcsec$ at
  5\,000~\kms, equivalent to 1.08~kpc diameter.
\end{minipage}

\end{table}

\begin{table}
  \caption{Summary of selection process.}
  \label{tab:selectpro}
  \begin{tabular}{lr}  \hline
    Selection step                        & \# of galaxies \\
                                          & remaining \\ \hline
    Original selection from FLASH survey  & 40 \\
    Morphology check with DSS             & 30 \\
    Observed galaxies                     & 24 \\
    Sufficiently high S/N                 & 22 \\
    Morphology check on auto-guider       & 17 \\
    FLASH redshifts confirmed             & 10 \\
    Morphology check with CCD imaging     &  9 \\ \hline
  \end{tabular}
\end{table}

\subsection{Central velocity dispersions}
\label{sec:dispersions}
Central velocity dispersion estimates were derived using version 8 of
the Fourier Quotient Coefficient algorithm developed by Bender (1990).
For this analysis the spectra were rebinned to a logarithmic wavelength
scale and a rest-wavelength range of approximately 4850 to 5560~\AA\/
was extracted. As we only consider central spectra in this paper we fit
a pure Gaussian profile to the broadening function, neglecting higher
order terms. For the final estimate of the central velocity dispersion,
we averaged the results obtained using 15 different template stars
ranging from G7III to K5III in spectral type.  The error was taken to
be the larger of (i) the mean internal error estimate and (ii) the rms
dispersion between the different template stars. The average relative
error is 3\%. The final, aperture-corrected (see
Appendix~\ref{sec:aperture}), central velocity dispersion measurements
are listed in Tables~\ref{tab:fieldsample} and \ref{tab:compsample}.

A comparison of central velocity dispersion measurements for the five
galaxies in common with Kuntschner (2000) shows excellent agreement;
the conversion factor between the two data-sets is $0.985\pm0.014$.
Since this is small and not significant, neither source is corrected
for the offset.

\section{Line-strength indices}
\label{sec:linestrengths}
The wavelength range of our observations allows us to analyse a range
of important absorption and emission features, including Balmer lines,
Mg and Fe lines and the \oii\/ and \oiiib\/ emission lines. These will
be compared to predictions from population synthesis models in order to
constrain parameters such as the luminosity-weighted age, metallicity
and magnesium to iron abundance ratio of each galaxy.

\subsection{Spectral resolution and choice of stellar population models}
In the optical region of the spectrum, the line-strength indices most
commonly exploited are those of the Lick/IDS system. The Lick index
definitions are described in detail in Worthey (1994), Worthey \&
Ottaviani (1997) and Trager et~al. (1998). Unfortunately, the Lick/IDS
system and its associated models (\eg\/ Worthey 1994; Vazdekis et~al.
1996) are calibrated at a fixed spectral resolution of $\sim$9~\AA\/
(FWHM). Thus to measure indices in the Lick system, it is necessary to
degrade the observed spectra to match this resolution, sacrificing the
weak features which may be most sensitive to the stellar populations.
The spectra of the LDR galaxies analysed in this paper were obtained at
an instrumental resolution of 2.6~\AA. For a comparison cluster sample,
we use the Fornax dataset of Kuntschner (2000); these spectra have an
instrumental resolution of 4.1~\AA. Hence, for all but the most
luminous galaxies (where velocity broadening unavoidably degrades the
resolution of the spectra), our measurements would be severely
compromised by the use of the Lick system as originally defined.

Instead, we will exploit the new stellar population models by Vazdekis
(1999), which predict the full SED of the integrated stellar light, at
the much higher resolution of 1.8~\AA\/ for two wavelength regions
(3856-4476~\AA\/ and 4795-5465~\AA). The Vazdekis models and their
application are discussed in greater detail in
Appendix~\ref{sec:vazdekis}. In order to make a fair comparison between
our LDR sample, the Fornax cluster sample, and the stellar population
models, we broadened the LDR data and the model SEDs to match the
4.1~\AA\/ resolution of the cluster sample. The predicted
line-strengths for simple stellar populations are obtained by measuring
the indices directly from the model SEDs.

\subsection{The measurements}
Although we measure and analyse the indices at higher resolution, we
continue to employ the index {\em definitions}\/ (\ie\/ the central and
pseudo-continuum bandpasses) standardized by the Lick group.

The measured indices were corrected for velocity dispersion broadening
using the results of simulations with template stars (see Kuntschner
2000 for details). Using the galaxies in common with the cluster sample
of Kuntschner (2000), we checked for any systematic offsets. There is
excellent agreement for all indices with formal offsets $<$0.1~\AA\/
($<$0.01 mag for Mg$_2$) which were removed from the Kuntschner (2000)
data if the offset was significant compared to its formal error. The
final line-strengths indices for the LDR sample are listed in
Table~\ref{tab:line_strengths}.

\begin{table*}
  \caption[]{Line-strength indices for galaxies in low-density regions}
  \label{tab:line_strengths}
  \begin{tabular}{lcccccccc} \hline
Name   &    \HdF       & \HgF          & \Hb           & Fe5015        & Mg$_2$          & \mgb          &  Fe5270       & Fe5335       \\
       &    [\AA]      &  [\AA]        &  [\AA]        & [\AA]         & [mag]           &  [\AA]        &  [\AA]        & [\AA]        \\ \hline
LDR\,08&$ 1.03\pm 0.14$&$-0.65\pm 0.13$&$ 1.23\pm 0.14$&$ 4.76\pm 0.29$&$ 0.190\pm 0.009$&$ 3.51\pm 0.14$&$ 2.87\pm 0.17$&$ 2.77\pm 0.19$\\
LDR\,09&$ 0.59\pm 0.09$&$-1.14\pm 0.08$&$ 1.91\pm 0.09$&$ 5.50\pm 0.19$&$ 0.214\pm 0.005$&$ 3.60\pm 0.09$&$ 3.30\pm 0.10$&$ 2.78\pm 0.12$\\
LDR\,14&$ 0.38\pm 0.15$&$-1.47\pm 0.14$&$ 2.05\pm 0.14$&$ 6.36\pm 0.31$&$ 0.231\pm 0.009$&$ 3.77\pm 0.15$&$ 2.99\pm 0.17$&$ 2.98\pm 0.20$\\
LDR\,19&$ 0.35\pm 0.19$&$-1.75\pm 0.18$&$ 1.30\pm 0.18$&$ 4.89\pm 0.41$&$ 0.256\pm 0.011$&$ 4.79\pm 0.19$&$ 2.77\pm 0.23$&$ 2.40\pm 0.28$\\
LDR\,20&$ 0.37\pm 0.14$&$-2.15\pm 0.14$&$ 1.13\pm 0.14$&$ 5.80\pm 0.32$&$ 0.274\pm 0.009$&$ 4.68\pm 0.15$&$ 3.30\pm 0.17$&$ 3.13\pm 0.21$\\
LDR\,22&$ 0.10\pm 0.10$&$-1.92\pm 0.10$&$ 1.47\pm 0.10$&$ 6.02\pm 0.27$&$ 0.288\pm 0.006$&$ 5.01\pm 0.12$&$ 3.20\pm 0.13$&$ 3.11\pm 0.17$\\
LDR\,29&$ 0.52\pm 0.14$&$-1.51\pm 0.13$&$ 2.27\pm 0.14$&$ 5.99\pm 0.30$&$ 0.248\pm 0.008$&$ 3.96\pm 0.14$&$ 3.45\pm 0.17$&$ 3.64\pm 0.20$\\
LDR\,33&$ 0.38\pm 0.17$&$-0.93\pm 0.16$&$ 2.11\pm 0.17$&$ 5.66\pm 0.38$&$ 0.229\pm 0.011$&$ 3.96\pm 0.18$&$ 3.29\pm 0.21$&$ 3.55\pm 0.25$\\
LDR\,34&$ 0.74\pm 0.12$&$-1.34\pm 0.11$&$ 1.83\pm 0.11$&$ 6.22\pm 0.24$&$ 0.227\pm 0.007$&$ 3.82\pm 0.11$&$ 3.11\pm 0.13$&$ 3.19\pm 0.16$\\ \hline
  \end{tabular}
  \begin{tabular}{lcc} \hline
Name   & \oii          & \oiiib           \\
       & [\AA]         & [\AA]            \\ \hline
LDR\,08&$-5.57\pm 0.36$&$-0.91\pm 0.10$   \\
LDR\,09&    $<0.5$     &     $<0.1$       \\
LDR\,14&    $<0.5$     &     $<0.1$       \\
LDR\,19&$-7.01\pm 0.54$&$-1.00\pm 0.13$   \\
LDR\,20&$-3.83\pm 0.39$&$-0.90\pm 0.10$   \\
LDR\,22&    $<0.5$     &     $<0.1$       \\
LDR\,29&    $<0.5$     &     $<0.1$       \\
LDR\,33&$-3.90\pm 0.44$&$-1.01\pm 0.12$   \\
LDR\,34&$-2.95\pm 0.29$&$-0.59\pm 0.07$   \\ \hline
  \end{tabular}

    \medskip
    \begin{minipage}{16.0cm}
      {\em Notes:}\/ All indices are calculated for the nominal
      aperture (equivalent to 1.08~kpc diameter) and are quoted without
      the correction for emission contamination. See
      Section~\ref{sec:linestrengths} for details.
    \end{minipage}
\end{table*}

\begin{figure}
\psfig{file=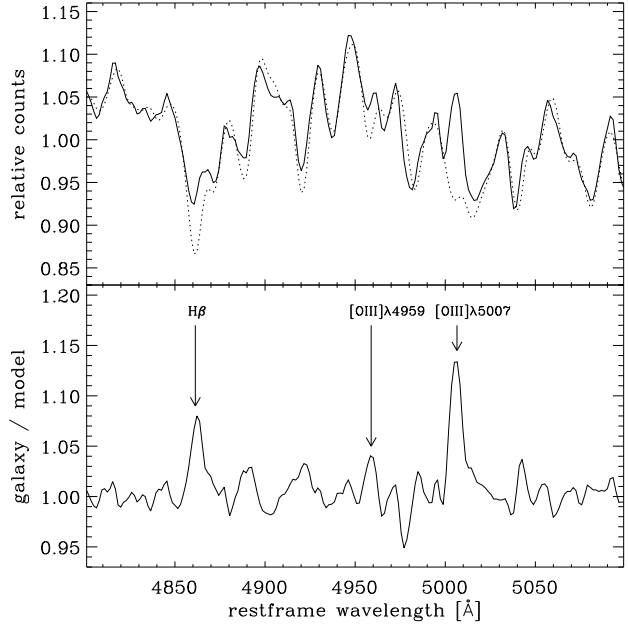,width=8.5cm}
\caption[]{\label{fig:oiii_emi}
  Measuring the \oiiib\/ emission. The top panel shows the spectrum of
  LDR\,19 (solid line) with the best-fitting Vazdekis model overlaid as
  the dotted line. The model is scaled to match the continuum level and
  broadened to the velocity dispersion of the galaxy; in this case the
  best-fit model has solar metallicity and 10~Gyr age. The \oiii\/
  emission lines at $\lambda = 4959,\, 5007$~\AA, as well as the
  emission contamination of the \Hb\/ absorption line, can be clearly
  seen. The ratio between observed spectrum and the stellar population
  model is shown in the bottom panel.}
\end{figure}

As we discuss in greater detail in the following section, several
indices can be contaminated by nebular emission lines. Corrections can
be established for this contamination, based on the measured equivalent
width of \oiiib. This line is difficult to discern in the spectra of
Figure~\ref{fig:fe_spec}. It can be reliably measured however, after
subtraction of a stellar absorption spectrum. We used a simple
minimization routine to find the best-fitting Vazdekis (1999) model,
over the range 4900--5100\AA, after broadening to match the velocity
dispersion of the galaxy. The \oiiib\/ equivalent width was then
measured from the residual spectrum, in which the \oiii\/ doublet is
clearly revealed (see Figure~\ref{fig:oiii_emi}).

\begin{figure}
\psfig{file=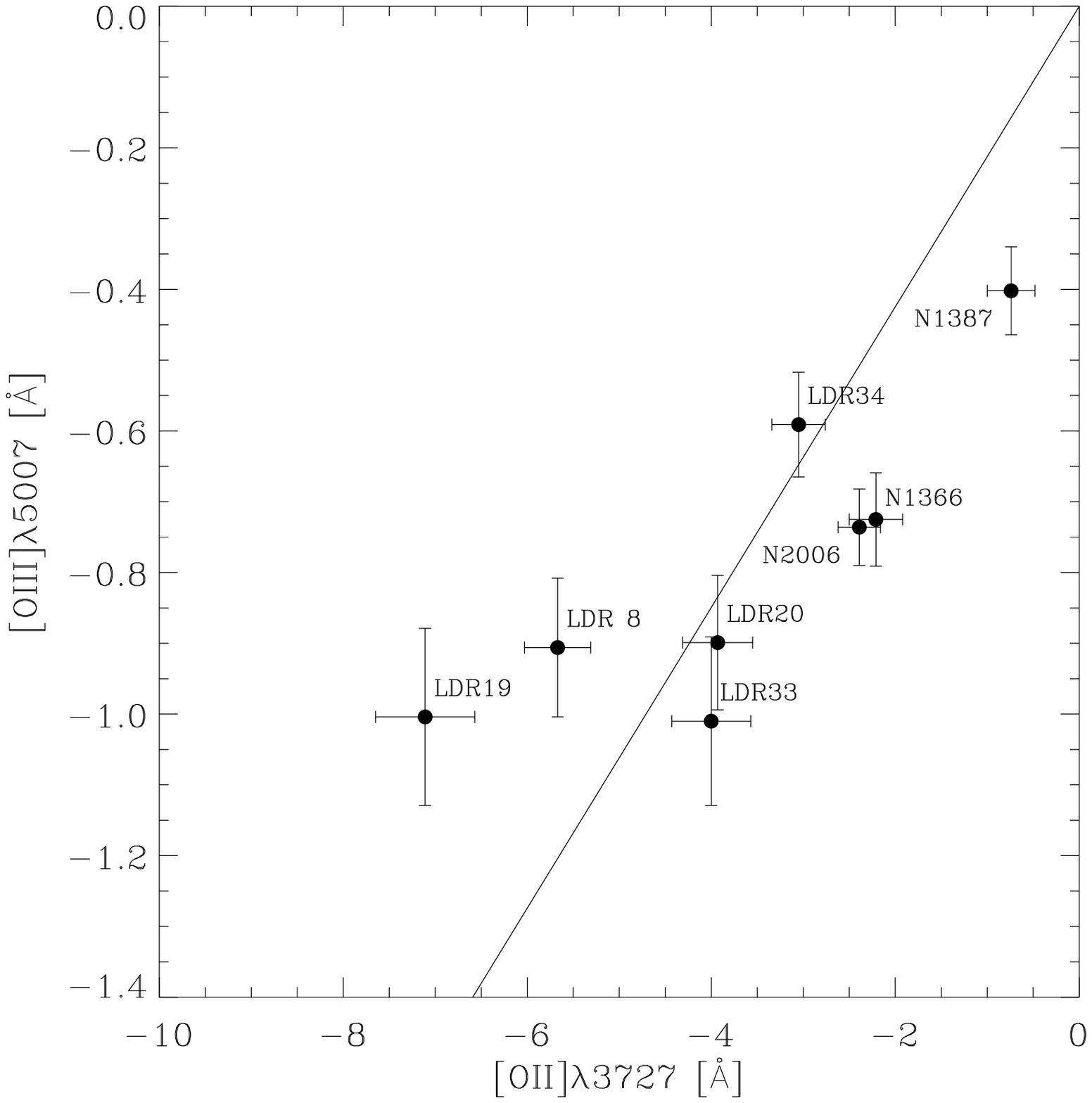,width=8.0cm}
\caption[]{\label{fig:emission}\oii\/ {\em vs}\/ \oiiib\/ emission
  diagram for galaxies with significant emission discussed in this
  paper. The best fitting linear relation (solid line) is
  \oiiib~$\simeq 0.21\, \times$\,\oii.}
\end{figure}

For completeness we show in Figure~\ref{fig:emission} the relation
between the \oii\/ and \oiiib\/ emission lines for all galaxies where
we could measure significant values. We determine a linear relation of
\oiiib~$\simeq 0.21\, \times$\,\oii. The equivalent width of the \oii\/
emission for all sample galaxies is $\le$7~\AA.

\subsection{Indices used in this paper}
For clarity, we summarize here the various absorption line indices
discussed in this paper, their sensitivities and their relative merits.

\subsubsection{Age-sensitive indices}
The features most useful for inferring ages in integrated spectra are
the Balmer series lines of Hydrogen. Given the spectral range of our
data, we employ \Hb\/ and \Hg\/ in this role.

A pitfall in using the Balmer lines is that the stellar {\it
  absorption}\/ features can be contaminated by nebular {\it emission}.
It is possible, in principle, to correct for this effect using an
assumed relationship between the Balmer emission and the strength of
the \oii\/ or \oiii\/ lines. For instance Trager et~al. (2000a)
concluded that 0.6 times the \oiiib\/ emission is a good estimate for
the \Hb-emission. However, in their sample of 27 galaxies the
correction factor varies from 0.33 to 1.25. Therefore it is doubtful
whether this correction is accurate for an individual galaxy (see also
Mehlert et~al. 2000), although it may be a good correction in a
statistical sense.

Much better established are the ratios between the Balmer
line-strengths in nebular emission spectra: \eg\/ \Hg/\Hb~$\approx
0.45$, \Hd/\Hb~$\approx 0.25$ (Osterbrock 1989). These ratios do not
carry through directly to the indices, because the Lick system does not
measure the equivalent width of the features alone, due to neighbouring
lines of other species in typical early-type galaxy spectra. However,
the ratios above can be used as input to simulations of nebular
contamination. Our tests with artificial contamination of galaxy
spectra give the following results: if the \Hb-index is contaminated
by, say, 1.0~\AA\/ emission then the \HgAF-indices will be affected by
$\approx0.6$~\AA\/ and the \HdAF-indices by only $\approx 0.4$~\AA.
Hence, using the results of Trager et~al. (2000a), the average
correction for the \HgAF\/ indices is $+0.36 \times |{\rmn [O{\small
    III}]} \lambda 5007|$, while for the \HdAF\/ indices it is $+0.22
\times |{\rmn [O{\small III}]} \lambda5007|$.

\begin{figure}
\psfig{file=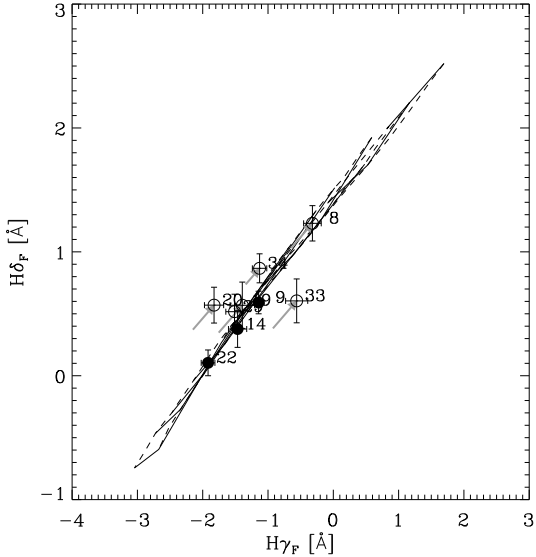,width=8.0cm}
\caption[]{\label{fig:hdelta}\HgF\/ {\em vs} \HdF\/ diagram for
  galaxies in low-density regions. Elliptical galaxies are represented
  by filled circles, S0s are shown as open circles. Data points are
  labelled with their LDR catalogue numbers (see
  Table~\ref{tab:fieldsample}). Grey arrows indicate the emission
  corrections. Overplotted are the predictions of stellar population
  models from Vazdekis (1999).}
\end{figure}

This alone would only give a factor of $\sim$1.7 improvement between
the \Hb-index and the \HgAF-indices, but one also has to take into
account the range spanned by the indices, compared to the error in
their determination. For example, at solar metallicity, the \Hb\/
absorption changes by 1.8~\AA\/ between a 17.8~Gyr model and a 1.6~Gyr
model, yet the \HgF-index changes by 4.4~\AA\/ for the same age range.
The total Balmer-index error is a function of S/N, but also dependent
on how well one can correct it for emission contamination. Hence the
optimal choice of the Balmer line depends on the dataset.  We will see
that the LDR galaxies exhibit nebular emission more frequently than is
typical for cluster E/S0s, and also with slightly greater strength.
Largely motivated by this fact, we will adopt the \HgF-index as our
age-sensitive index, since it significantly reduces the emission
contamination compared to the \Hb-index, while it shows similar error
statistics. Although \Hd\/ is available for the LDR sample, it is not
covered by the Fornax data. For completeness we show the relation
between the \HgF\/ index and the \HdF\/ index for the galaxies in
low-density regions in Figure~\ref{fig:hdelta}.

Abundance ratio influences on the \Hb-index are thought to be small
\cite{tra00a} and should therefore not compromize our results. The
dependence of the \HgF-index on abundance ratios is unknown and will
not be considered in this paper.

\subsubsection{Metallicity-sensitive indices}
The metallicity-dependent indices in the spectral region observed are
associated principally with iron and magnesium dominated features.
Since early-type galaxies generally exhibit non-solar abundance ratios
of Mg relative to Fe \cite{pel89,wor92,wei95,jor99,tra00b,kun00} and
probably also other elements (\eg\/ Vazdekis et~al. 1997), metallicity
estimates are dependent on which index or combination of indices is
employed.

In order to suppress these metallicity biases, and to compare with
theoretical model prediction of hierarchical galaxy formation models,
we aim to measure a mean metallicity representing all elements.
Observationally this is very difficult and can only be approximated.
For example, \gon\/ defined the [MgFe]-index\footnote{${\rmn [MgFe]} =
  \sqrt{{\rmn Mg}\,b \times ({\rmn Fe5270} + {\rmn Fe5335})/2}$\,,
  \cite{gon93}} which combines the $\alpha$-element Mg with the
contributions of Fe. Although the exact combination and relative
abundances of $\alpha$-elements to Fe in early-type galaxies are not
well determined (\eg\/ Vazdekis et al.  1997, Worthey 1998, Vazdekis
et~al. 2001), the [MgFe] index provides a good first-order estimate of
the mean metallicity \cite{tra00a,kun01}.

\begin{figure}
\psfig{file=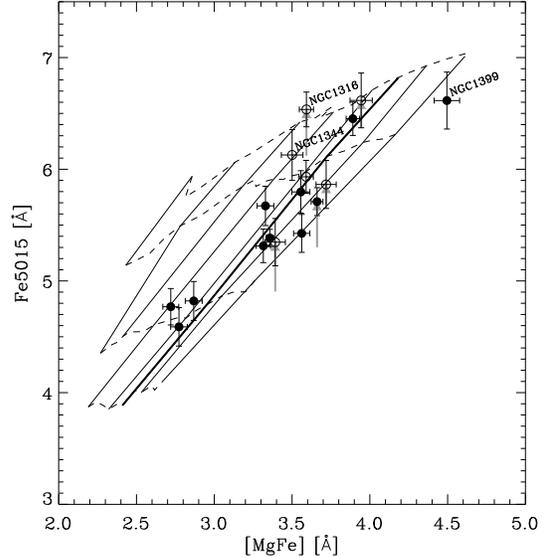,width=8.0cm}
\caption[]{\label{fig:mgfefe5015}The relation between the [MgFe] and
  Fe5015 indices for Fornax cluster galaxies. Elliptical galaxies are
  represented by filled circles, S0s are shown as open circles.  The
  Fe5015 index has been corrected for \oiii-emission contamination
  (grey arrows). Overplotted are predictions of stellar population
  models by Vazdekis (1999). The thick solid line indicates the
  predictions for a constant age of 10~Gyr over a metallicity range of
  ${\rmn [M/H]} = -0.7$ to +0.2.}
\end{figure}

While the [MgFe]-index is a good mean metallicity indicator it also
shows a significant dependence on age variations as we will see in
Section~\ref{sec:ages}. An index which is less dependent on age, while
similarly being sensitive to a wide range of elements, is Fe5015. We
show in Figure~\ref{fig:mgfefe5015} the relation between [MgFe] and
Fe5015.  The indices show a good correlation and we conclude that
Fe5015 can also be used as a good mean metallicity indicator.

The Fe5015 index is affected by \oiiib\/ emission in its central
bandpass, and by \oiiia\/ emission in its blue continuum bandpass. By
artificially adding \oiiia\/ and \oiiib\/ emission to galaxy spectra
and examining the effects on the Fe5015 index, we have established that
the emission contamination can be corrected by adding $+0.61(\pm0.01)
\times |{\rmn [O{\small III}]} \lambda 5007|$ to the Fe5015 index. It
is important to note that the error of this correction is dominated by
the measurement error of \oiiib\/ and not by the error in the
correction factor of 0.61 (see also Figure~\ref{fig:oiii_emi}). This is
in contrast to the Balmer line correction based on \oiiib, where the
correction factor dominates the error since the true ratio varies
greatly from galaxy to galaxy \cite{tra00a}.

Note that the C$_2$4668-index, which is perhaps also a good indicator
of mean metallicity \cite{wor94a,kun98}, is not used here, since the
Jones (1999) library on which the Vazdekis (1999) models are based do
not cover this region of the spectrum.

\section{RESULTS}
\label{sec:results}
In our analysis of early-type galaxies in low-density regions we focus
on three key stellar population diagnostics: (i) the Mg--$\sigma$
relation, (ii) the luminosity-weighted ages and metallicities as
inferred from line-strength diagrams and (iii) the abundance ratios as
determined from Mg and Fe absorption lines. In this analysis we take a
two-fold approach by comparing the observed quantities directly with a
sample of early-type galaxies in the Fornax cluster and by analysing
the observed quantities with respect to stellar population models.

Most of the Fornax cluster data was published in Kuntschner \& Davies
(1998) and Kuntschner (2000). However, we have re-observed four
galaxies during this project and also added a further three early-type
galaxies in the Fornax cluster (see Table~\ref{tab:compsample}). In
order to make a fair comparison with the LDR sample we only select
galaxies in the Fornax cluster with a central velocity dispersions of
$\sigma > 75$~\kms, yielding a sample of 17 cluster galaxies (11 Es and
6 S0s). The cut in velocity dispersion was chosen to match the faint
end luminosity and mass distribution (as measured by the central
velocity dispersion) of our sample of galaxies in low-density regions
(see also Figure~\ref{fig:selection}). We did not attempt to match the
mass distribution for the bright galaxies since we did not select
against these galaxies in the LDR sample. A velocity dispersion cut
would remove only two galaxies from the Fornax sample (NGC\,1399 and
NGC\,1404). The results presented in this section do not depend
critically on the details of selecting the comparison sample; the small
effects of matching cluster and field luminosity distributions are
commented on, where relevant. We emphasize that in this study we
re-analyse the Fornax sample at 4.1~\AA\/ (FWHM) spectral resolution
and also correct the data to the nominal aperture of the LDR sample
(see Appendix~\ref{sec:aperture}).

\subsection{Mg$_2$--$\sigma$ relation}
Early-type galaxies in clusters show a tight relation between the
absorption strength of the Mg feature at $\sim$5175~\AA\/ and their
central velocity dispersion. This relation has been used to probe
galaxy formation nearby (\eg\/ Colless et~al. 1999, Kuntschner et~al.
2001) and at medium redshift (\eg\/ Ziegler \& Bender 1997; Ziegler
et~al. 2001). Although the interpretation of the Mg--$\sigma$ relation
is complicated by the degenerate effects that particular combinations
of age and metallicity can have on the Mg line-strength, these studies
have generally concluded that most of the stars in cluster E/S0s formed
at $z \ge 2$.

\begin{figure*}
\psfig{file=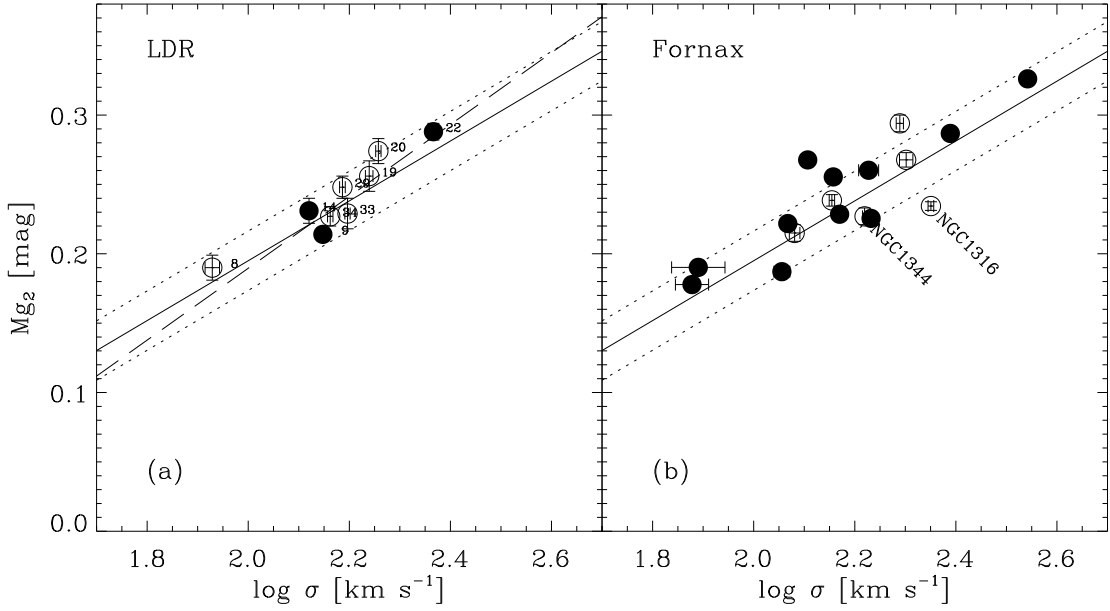,width=16.0cm}
\caption[]{\label{fig:mg_sig}The Mg$_2$--$\sigma$ relation for the
  LDR galaxies (panel~a, data points are labelled with their catalogue
  numbers) and the comparison Fornax cluster sample (panel~b).
  Elliptical galaxies are represented by filled circles, S0s are shown
  as open circles. The long dashed line shows a straight-line fit to
  the LDR sample, taking into account errors in x- and y-direction. The
  solid line in both panels represents a similar fit to the Fornax galaxies,
  while  the dotted lines show the $1 \sigma$ spread in the Fornax relation.}
\end{figure*}

The galaxies in the Fornax cluster show a tight Mg$_2$--$\sigma$
relation with an observed rms scatter of only 0.021~mag (see
Figure~\ref{fig:mg_sig}b):

\begin{equation}
  \label{eq:metal_sig2}
  {\rmn Mg} _2 = 0.216(\pm 0.025) \log \sigma - 0.236 (\pm0.057) \, .
\end{equation}
Note that the observed scatter is much larger than the typical
observational error: the dominant source of dispersion is the intrinsic
scatter in properties from galaxy to galaxy.

The observed scatter and slope are in good agreement with the results
from recent studies of larger samples of mostly cluster early-type
galaxies \cite{jor97,coll99,kun01}. NGC\,1316 (Fornax~A), the prominent
merger galaxy in the Fornax cluster (Schweizer 1980, 1981), shows weaker
than average Mg$_2$ absorption for its velocity dispersion ($\Delta
\rmn{Mg}_2 = -0.036\, \rmn{mag}$) consistent with the effects of young
stellar populations. However, it is surprising that this merger galaxy
doesn't deviate more clearly from the mean relation as recent estimates
of its luminosity-weighted mean age are around 2--3~Gyr
\cite{kun98,mac98,gom01,gou01}. Stellar population models \cite{vaz99}
predict $\Delta \rmn{Mg}_2 = -0.074\, \rmn{mag}$ for a change in
luminosity-weighted age from 10~Gyr to 2.5~Gyr at solar metallicity.

The galaxies in low-density regions (Figure~\ref{fig:mg_sig}a) follow
an Mg$_2$--$\sigma$ relation very similar to that of the cluster
sample. A formal straight line fit results in a slightly steeper
relation (long dashed line). However, due to the small numbers of
galaxies this difference in fitted slope is not significant.  A fit to
the LDR galaxies with a fixed slope of 0.216 (as found for the Fornax
cluster) gives a difference (${\rmn LDR} - {\rmn Fornax})$ of
0.004$\pm$0.003 mag in the intercept, which is not significant. The
observed rms scatter with respect to the slope of the Fornax galaxies
is 0.012~mag. Surprisingly, this is significantly smaller than the
scatter measured for the Fornax cluster (F-test significance $>90\%$).
However, when we use the \mgb--$\sigma$ relation instead of the
Mg$_2$--$\sigma$ relation, we find an observed scatter of the same size
for the LDR sample as for the Fornax cluster.

In our small sample of galaxies we find no significant shift in the
zero-point between the cluster galaxies and the LDR galaxies.
Furthermore there is no evidence for {\it increased} scatter in the LDR
sample, which might be expected if these have experienced more diverse
formation histories. Taken at face value, the Mg--$\sigma$ relation
would suggest that the star-formation history of LDR and cluster
galaxies is very similar. This test on its own, however, is not a
conclusive test of the star-formation histories since an
anti-correlation of age and metallicity effects can conspire to produce
a tight Mg--$\sigma$ relation while hiding a complex star-formation
history \cite{coll99,tra00b,kun01}. NGC\,1316 in the Fornax cluster is
perhaps a good example of this scenario.

In the next section we will explore the luminosity-weighted ages and
metallicities in a more direct approach.

\subsection{Ages and metallicities}
\label{sec:ages}
Luminosity-weighted ages can be inferred from an age/metallicity
diagnostic diagram by plotting an age-sensitive index and a
metallicity-sensitive index against each other. In order to make age
and metallicity estimates, we use the Vazdekis (1999) models. We
describe these models and their differences with respect to other
models in Appendix~\ref{sec:vazdekis}. In this section, we will present
first a diagram of the [MgFe] index {\em vs}\/ \Hb\/ in order to
demonstrate the nebular emission contamination which affects this
combination, especially for our LDR sample. Then, by using a higher
order Balmer line (\HgF), we will analyse diagrams with greater
robustness against nebular emission.

\begin{figure*}
\psfig{file=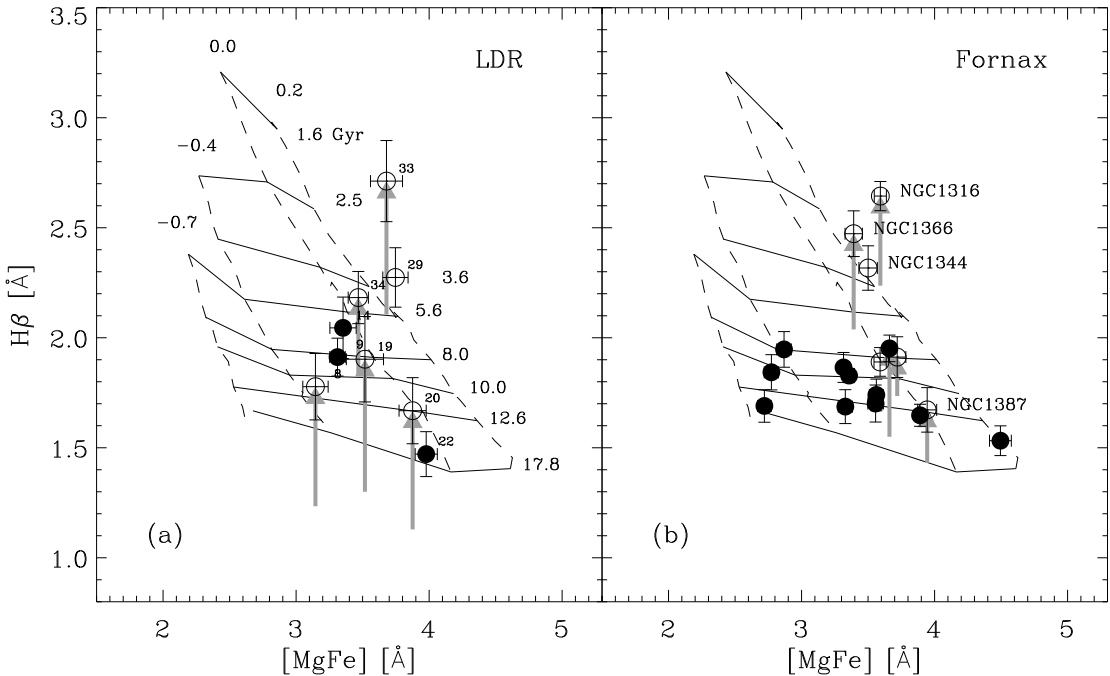, width=16.0cm}
\caption[]{\label{fig:mgfe_hb}\Hb\/ {\em vs}\/ [MgFe] diagram for the
  LDR galaxies (panel~a) and the Fornax cluster sample (panel~b).
  Galaxies in low-density regions are labelled with their catalogue
  numbers. Elliptical galaxies are represented by filled circles, S0s
  are shown as open circles. Grey arrows indicate an emission
  correction for the \Hb-index. Overplotted are the predictions of
  stellar population models from Vazdekis (1999) at a spectral
  resolution of 4.1~\AA\/ (FWHM). The solid lines are lines of constant
  age and the dashed lines are lines of constant metallicity. The
  models span an age range of 1.6 to 17.8~Gyr, the metallicity range is
  $-0.7 < {\rmn [M/H]} < +0.2$. The age and metallicity steps are
  indicated at the top and to the right of the model predictions.
  Throughout this paper we will use for the stellar population models
  the spectral resolution, and the age \& metallicity range as given
  above.}
\end{figure*}

Figure~\ref{fig:mgfe_hb} shows a diagram of [MgFe] against \Hb\/ for
the LDR sample as well as the Fornax cluster galaxies. The emission
corrections for the \Hb\/ index are indicated by grey arrows. Prior to
any emission correction (see Section~\ref{sec:linestrengths}), the LDR
sample spans a wide range in \Hb\/ line-strength from 1.1 to 2.3~\AA.
However, the \Hb\/ line-strength measurements below 1.4~\AA\/ are
strongly affected by emission. After the emission correction, affecting
five out of nine galaxies, all \Hb\/ measurements are larger than
$\sim$1.4~\AA, which is consistent with the model predictions by
Vazdekis (overplotted on Figure~\ref{fig:mgfe_hb}).  Although the
corrected values indicate a large luminosity-weighted age range from
2.5 to 18~Gyr it is difficult to draw any firm conclusions from this
diagram due to the uncertainties in the emission correction, which can
be as large as the correction itself for the \Hb-index. The
distribution of galaxies in the \Hb\/ {\em vs} [MgFe] diagram is
reminiscent of the shell and pair galaxies observed by Longhetti et~al.
(2000).

The Fornax sample of galaxies (Figure~\ref{fig:mgfe_hb}b) contains only
five galaxies out of 17 with emission corrections. Furthermore the
correction is, on average, slightly smaller than that of the galaxies
in low-density regions. Therefore this diagram is much more useful to
infer luminosity-weighted ages and metallicities for Fornax galaxies,
than its counterpart for the LDR sample. In Fornax, we find mainly a
sequence of metallicity at age $\sim$12~Gyr, with only three galaxies
having clearly stronger \Hb\/ line-strength and therefore younger ages
(NGC\,1316, NGC\,1344 and NGC\,1366). This re-analysis of the Fornax
data is in excellent agreement with the earlier study by Kuntschner
(2000).

We find that our LDR sample shows a marginally higher fraction of
galaxies ($0.56 \pm 0.17$) with significant emission lines (\eg\/
\oii\/ and \oiiib) than our cluster sample ($0.29 \pm 0.11$). When
\oiiib\/ emission is present it also tends to be slightly stronger for
these galaxies (Fornax: $-0.55\pm0.19$~\AA, Field: $-0.88\pm0.17$~\AA).
After the emission corrections, the galaxies in low-density regions
show on average stronger \Hb\/ absorption than the ellipticals in the
Fornax cluster. We find approximately three galaxies which occupy the
same region in the diagram as NGC\,1316 and NGC\,1344, which both show
signs of a recent merger.

\begin{figure*}
\psfig{file=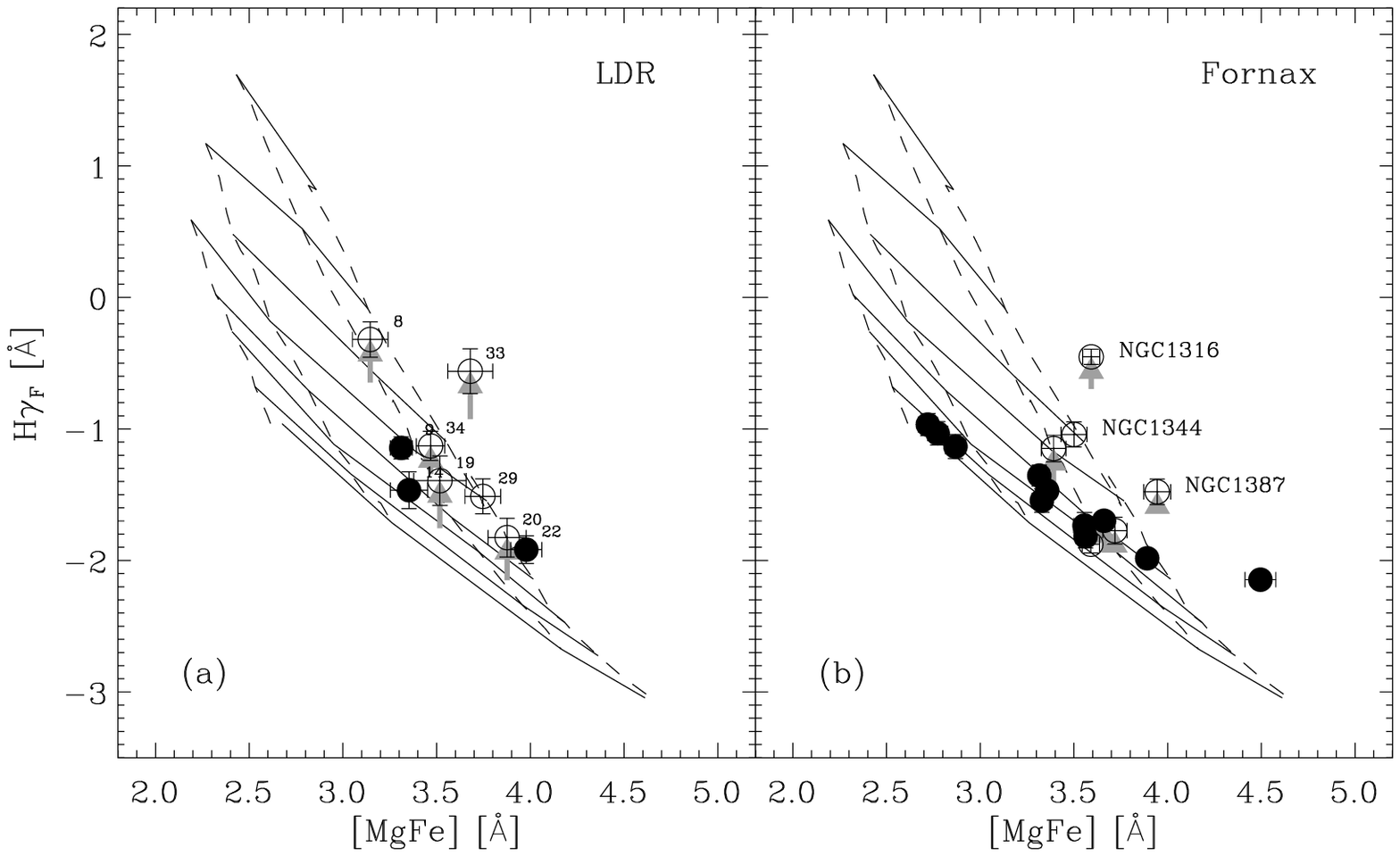,width=16.0cm}
\psfig{file=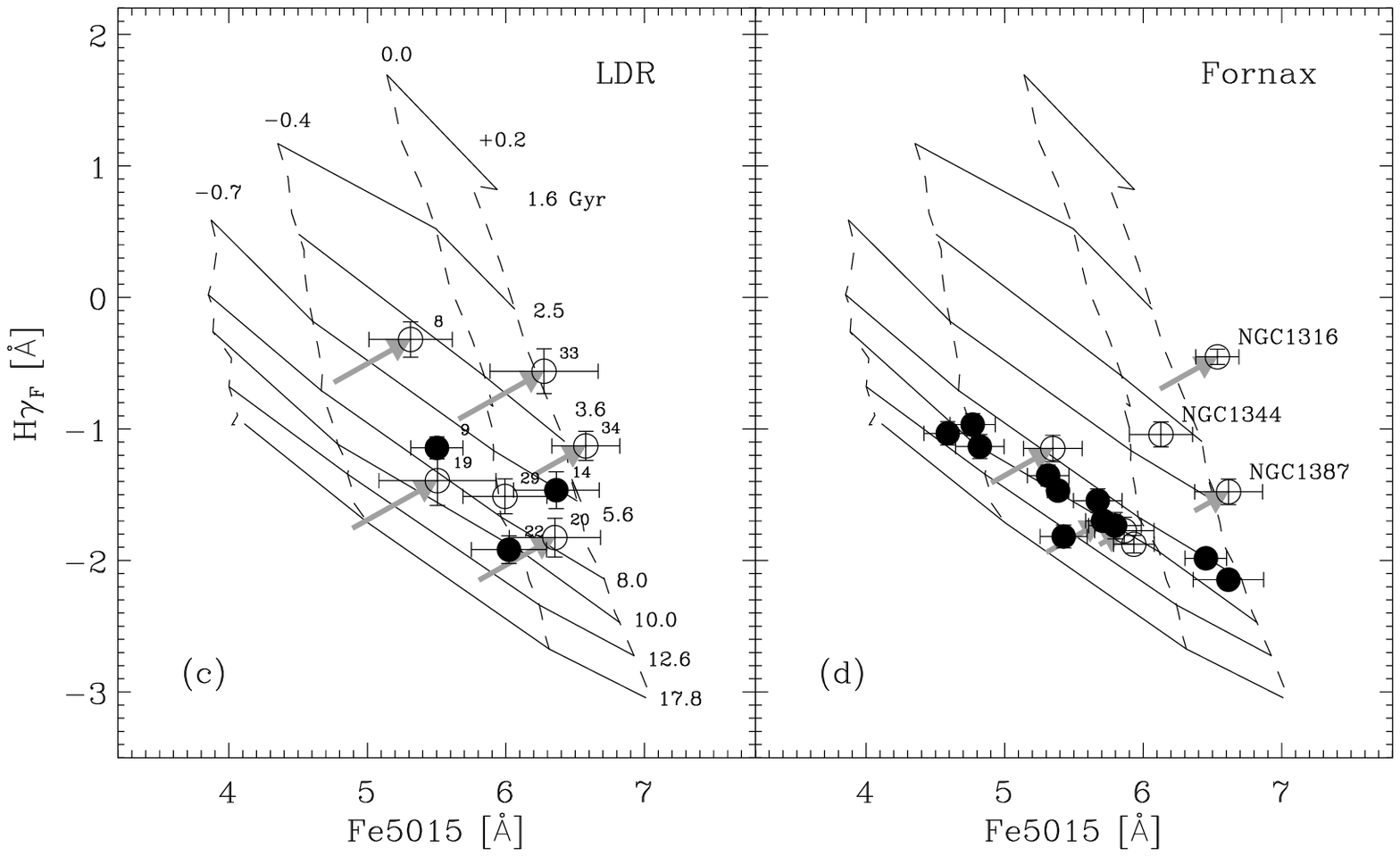,width=16.0cm}
\caption[]{\label{fig:mgfe_hgf}Panels (a) \& (b): \HgF\/ {\em vs} [MgFe]
  diagram for galaxies in low-density regions (left hand panel) and the
  Fornax cluster sample (right hand panel). Galaxies in low-density
  regions are labelled with their catalogue numbers. Elliptical
  galaxies are represented by filled circles, S0s are shown as open
  circles. Grey arrows indicate an emission correction for the \HgF\/
  index. Note that the emission corrections are much smaller than those
  for \Hb\/ (Figure~\ref{fig:mgfe_hb}).  Panels (c) \& (d): \HgF\/ {\em
    vs}\/ Fe5015 diagram. Both \HgF\/ and Fe5015\/ have been corrected
  for emission contamination, as indicated by the grey arrows (see
  Section~\ref{sec:ages} for details). Overplotted are the predictions
  of stellar population models from Vazdekis (1999), as in the previous
  figure. The individual steps in age and metalicity are indicated in
  panel (c).}
\end{figure*}

Figures~\ref{fig:mgfe_hgf}a \& b show the [MgFe] {\em vs}\/ \HgF\/
diagram for the LDR sample and the Fornax cluster galaxies. The reduced
sensitivity against Balmer emission can be clearly seen in the reduced
size of the emission corrections (grey arrows). By using the \HgF\/
index as age indicator we are less sensitive to emission, but at the
same time the sensitivity to age is reduced compared to the \Hb-index.
This can be seen in Figure~\ref{fig:mgfe_hgf} where the lines of
constant age and constant metallicity, as predicted by the stellar
populations models, are not as well separated as in
Figure~\ref{fig:mgfe_hb}. Again, the Fornax cluster shows a tight
sequence of galaxies along roughly a constant age line with only three
lenticular galaxies scattering towards younger luminosity-weighted
ages. The LDR galaxies span a wide range in ages, from as old as the
Fornax ellipticals to as young as NGC\,1316.

In order to improve the separation of age and metallicity effects we
plot in Figure~\ref{fig:mgfe_hgf}c \& d the \HgF\/ {\em vs}\/ Fe5015
diagram. Relative to [MgFe], the Fe5015 index shows reduced sensitivity
to changes in the age of a stellar population. Clearly this new index
combination disentangles the effects of age and metallicity more
powerfully than the previous diagram. Emission corrections, which
affect both indices, have been applied to the indices (grey arrows).

From the \HgF\/ {\em vs}\/ Fe5015 diagram we measured the ages and
metallicities with respect to the Vazdekis (1999) models. For this
purpose, we linearly interpolated between the model grid points and
also linearly extrapolated the models to a metallicity of ${\rmn [M/H]}
= 0.4$ (necessary for two galaxies in each sample).  We note that the
resulting ages and metallicities are subject to systematic offsets. For
example, had we used a different index combination, we would have
determined slightly different {\em absolute}\/ ages and metallicities.
However, the relative ranking within the sample and between the samples
would not be significantly different. For instance, the ages measured
from the [MgFe]--\Hb\/ diagram are on average $\sim$1~Gyr older than
those obtained from the Fe5015--\HgF\/ analysis. The metallicities show
a negligible offset. These comparisons exclude galaxies with
significant emission.

\begin{figure*}
  \psfig{file=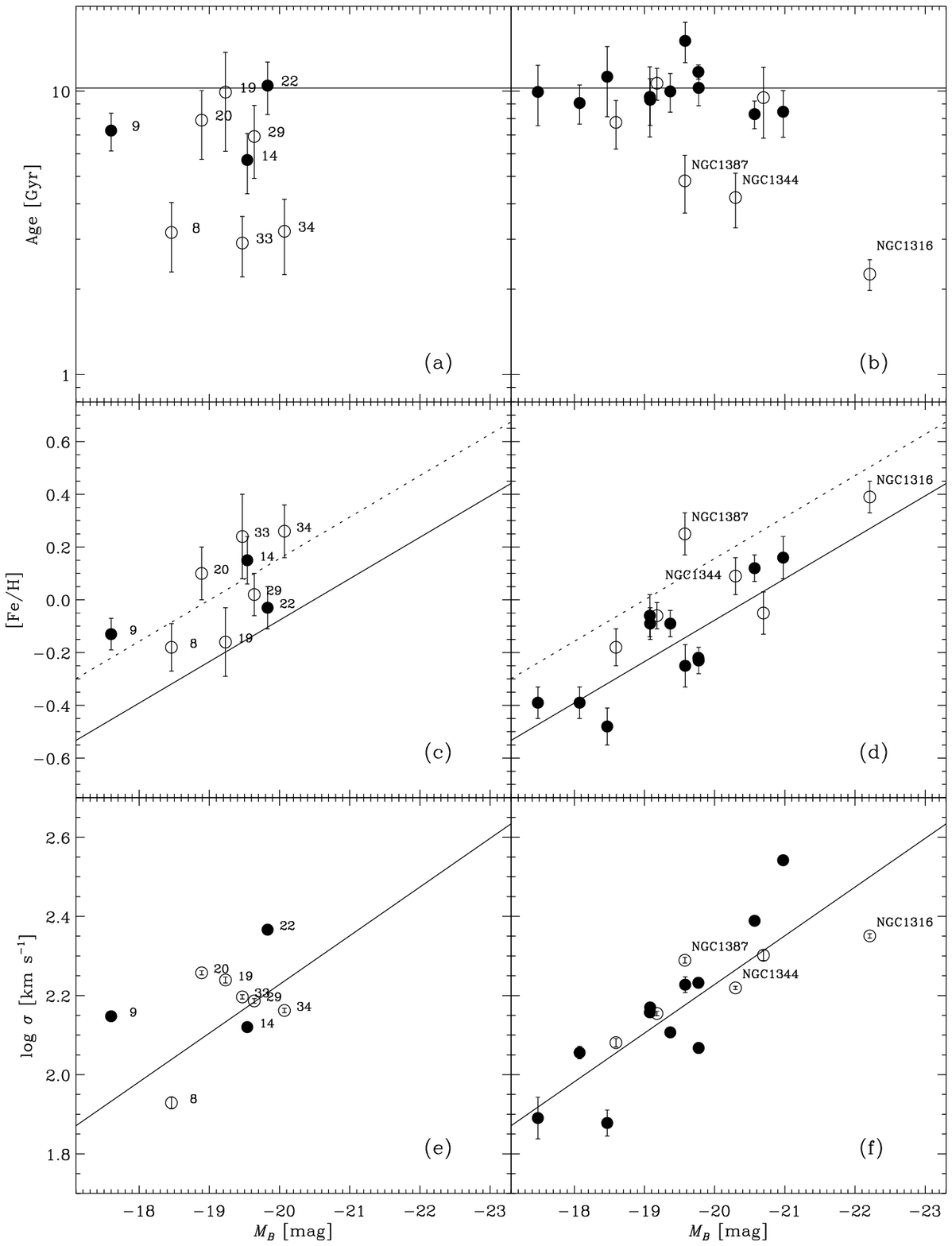,width=16.0cm}
\caption[]{\label{fig:age_metal_MB}Luminosity-weighted ages,
  metallicities and central velocity dispersions are plotted as a
  function of $M_B$ for the LDR sample (panels a, c, e) and the Fornax
  cluster (panels b, d, f). Galaxies in low-density regions are
  labelled with their catalogue numbers.  The ages and metallicities
  were measured from the \HgF\/ {\em vs}\/ Fe5015 diagram (see
  Figure~\ref{fig:mgfe_hgf}c, d) with respect to the Vazdekis (1999)
  models. For this purpose the Vazdekis models were linearly
  extrapolated to a metallicity of ${\rmn [M/H]} = +0.4$.  Errors on
  the age and metallicity estimates were derived by adding and
  subtracting the index error for each galaxy individually and
  re-deriving the age/metallicity estimates, with the final uncertainty
  taken to be 0.7 times the maximum change in metallicity and age. The
  errors on age and metallicity are correlated; for details see
  Kuntschner et~al. (2001). The dashed line in panels a \& b indicates
  the error-weighted mean age of the Fornax ellipticals.  The solid
  lines in panels c to f represent a linear fit to the
  metallicity--$M_B$ and Faber-Jackson relation of the Fornax cluster
  galaxies. The dotted line in panels c \& d show a fit to the LDR
  sample, with the slope fixed to that of the Fornax relation.}
\end{figure*}

In Figure~\ref{fig:age_metal_MB}, we show the estimated ages and
metallicities as a function of absolute $B$-magnitude for both the LDR
sample and the Fornax cluster sample. The Fornax cluster shows a tight
sequence of elliptical galaxies with an average age of 10.2~Gyr
(indicated by the dashed line in Figure~\ref{fig:age_metal_MB}b). The
S0s show a large range in ages. The average age of elliptical and
lenticular galaxies in low-density regions ($6.4\pm1.0$~Gyr) is lower
compared to the E/S0s in the Fornax cluster ($8.9\pm0.7$~Gyr; errors
are quoted as errors on the mean). If we select our comparison sample
by applying an upper mass cut (velocity dispersion), the average age of
Fornax galaxies is $8.5\pm0.7$~Gyr. Matching the luminosity range for
the bright end of field and cluster samples gives for the Fornax
cluster a mean age of $9.9\pm0.7$~Gyr.  Thus, our result is not
sensitive to the detailed selection of the comparison sample.
  
The age difference between cluster and low-density region sample can
perhaps be partly explained by the higher relative fraction of
lenticular galaxies in the low-density region. Unfortunately, given the
very small sample size, it is not possible to determine whether
environmental effects act to change the properties of certain
subclasses of galaxies (e.g. S0s and ellipticals), or whether the
overall effect is due to a change in the relative proportions of such
subclasses in the population. It is quite possible that both mechanisms
apply, and the distinction between them is not necessarily entirely
clear.

For both the Fornax and LDR samples, our metallicity measurements show
a correlation with absolute $B$-magnitude in the sense that brighter
galaxies are more metal rich. We note that the LDR sample does not
contain any low metallicity ([Fe/H]$\simeq$-0.4) galaxies even at the
faint end of our sample, whereas the Fornax cluster shows three
elliptical galaxies in this metallicity range. In fact, at any given
luminosity, the LDR sample shows on average a stronger metal content
than the cluster sample ($\Delta\, \rmn{[Fe/H]} \simeq
0.23\pm0.03$~dex, see Figure~\ref{fig:age_metal_MB}c, dotted line). The
last statement is also true if we use the central velocity dispersion
as x-axis in Figure~\ref{fig:age_metal_MB}, albeit with a smaller
metallicity offset of $\Delta\, \rmn{[Fe/H]} \simeq 0.15\pm0.03$~dex.

The random and systematic errors affecting the age and metallicity
estimates are correlated, such that an overestimation of the age leads
to an overestimation of the metal content. Is it conceivable, then,
that a systematic effect is responsible for the difference between LDR
and Fornax samples in Figure~\ref{fig:age_metal_MB}? The analysis of
the two samples is identical, with the same spectral resolution,
stellar population models and measurement techniques, therefore
minimizing systematic offsets between the samples. Clearly a potential
source of error lies in the emission corrections, which are slightly
larger and more frequent for the LDR galaxies than for the Fornax
sample. Our principal results are, however, quite robust against the
choice of index combination used for the age and metallicity estimates
(\ie\/ the [MgFe]--\Hb\/ and Fe5015--\HgF\/ diagrams yield the same
conclusions). We are therefore confident that the differences between
the two samples in Figure~\ref{fig:age_metal_MB} do indeed indicate
that E/S0s in low-density regions harbour younger and more metal-rich
stellar populations than cluster  E/S0s of comparable luminosity.

\subsection{Abundance ratios}
In this section, we investigate the Mg-to-Fe abundance ratios in our
sample of LDR galaxies. While, as shown in the previous section, one
can obtain estimates of the luminosity-weighted ages and metallicities
from line-strength indices, it is more difficult to extract information
on the duration and strength of the star-formation process. The
abundance ratios of certain elements, \eg\/ [Mg/Fe], however, can carry
crucial information about the star-formation time-scales.

The chemical enrichment process in galaxies is predominantly driven by
the ejecta of SN~Ia (the main producer of Fe peak elements) and SN~II
(producing mainly alpha elements such as Mg). Because SN~Ia
($\sim$1~Gyr time-scale) are delayed compared to SN~II which explode on
short time-scales ($\le10^6-10^7$ yr), the [Mg/Fe] ratio is determined
by (i) the duration of the star-formation and (ii) the initial mass
function (IMF, see \eg\/ Worthey, Faber \& Gonz\'{a}lez 1992).

It is now well established that elliptical galaxies in {\em clusters}
show non-solar abundance ratios of certain elements. For example
[Mg/Fe] is generally positive in the range 0.1 to 0.3
\cite{pel89,wor92,wei95,jor99,tra00b,kun00,kun01}. This is generally
interpreted as evidence for a short star-formation time-scale (perhaps
combined with a top-heavy IMF) in early-type galaxies, \ie\/ the star
formation stopped before the products of SN~Ia could be incorporated
into the stars we observe today. In environments where star formation
can continue for a longer time, \eg\/ the disc of our own galaxy, one
expects solar abundance ratios for the younger, more metal rich stars,
which indeed is the case \cite{edv93,mcw97}.

Following Trager et~al. (2000a), we estimate the [Mg/Fe] ratios with the
help of a \fe\/ vs \mgb\/ diagram and stellar population models. For
the detailed procedure see Kuntschner et~al. (2001). The
luminosity-weighted ages, which are needed to make the [Mg/Fe]
estimates, were taken from the \HgF\/ {\em vs}\/ Fe5015 diagram. We note
that for recent (\ie\/ $\le$1~Gyr) star-bursts contributing only a
small fraction in mass of the total galaxy, the above method of
estimating the abundance ratios can lead to an overestimation of
[Mg/Fe] \cite{kun00}. Hence the abundance ratio measurements for the
younger galaxies should be interpreted somewhat more cautiously.

\begin{figure*}
  \psfig{file=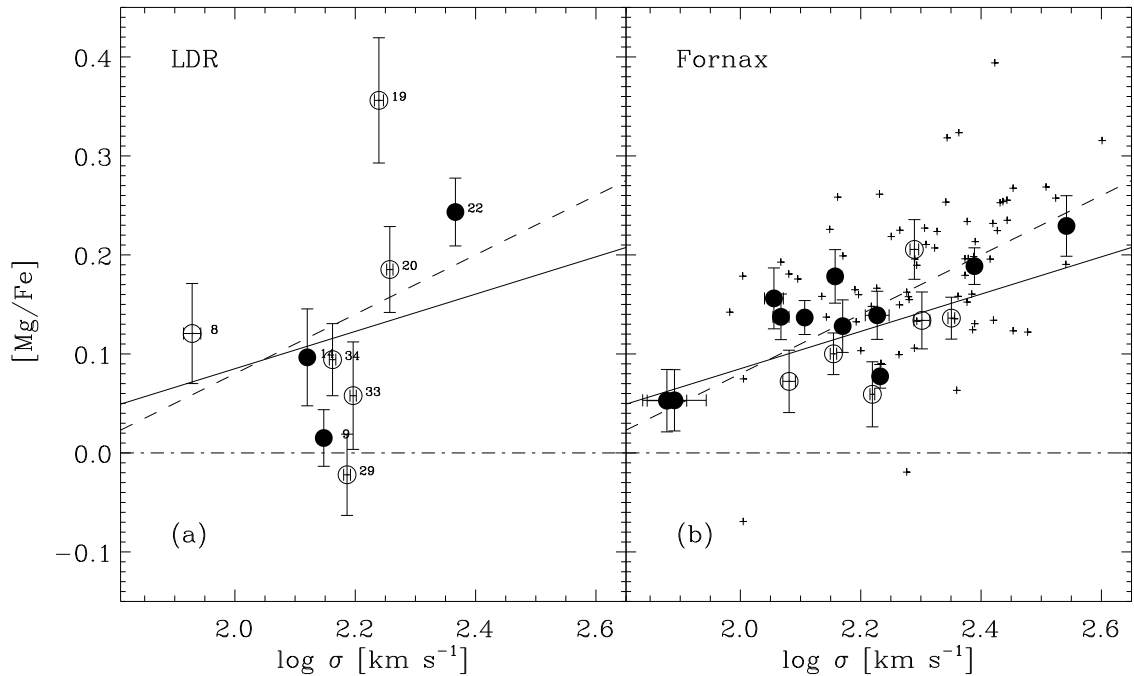,width=16.0cm}
\caption[]{\label{fig:abundance}[Mg/Fe] {\em vs}\/ $\log \sigma$
  relation for our sample in low-density regions (panel~a) and for the
  Fornax cluster sample (panel~b). The LDR galaxies are
  labelled with their catalogue numbers. The solid line shows a
  straight-line fit to the Fornax data, taking into account errors in
  both variables. The dashed line represents the best-fitting relation
  to the sample of Kuntschner et~al. (2001, plus signs). The latter
  sample  is dominated by early-type galaxies in the Coma
  and Virgo clusters. The dot-dashed line marks the region of solar
  abundance ratios.}
\end{figure*}

Our estimates of [Mg/Fe] are shown in Figure~\ref{fig:abundance}, as a
function of central velocity dispersion. In the Fornax cluster, most of
the galaxies have ratios of [Mg/Fe] = 0.1 to 0.2, with a weak tendency
towards larger [Mg/Fe] ratios for more massive galaxies. A
straight-line fit, taking errors in both variables into account, is
shown as solid the line. The relation found by Kuntschner et~al. (2001)
for a larger sample of mostly cluster elliptical galaxies, shown as
plus signs in Figure~\ref{fig:abundance}, is slightly steeper (dashed
line).  However, the [Mg/Fe] values for Fornax are in good agreement
with the Kuntschner et~al. (2001) sample. Note that the Kuntschner
et~al. (2001) data was not corrected to the nominal aperture which we
use in this study.  However, preliminary investigations into the radial
gradients of [Mg/Fe] in elliptical galaxies show that the abundance
ratios are roughly constant at least out to one effective radius
\cite{kun98b,hall99,dav01}.

For the galaxies in our LDR sample, we estimate a range in ${\rmn
  [Mg/Fe]} \simeq 0.0$ to $\sim$0.3. These values are similar to the
[Mg/Fe] ratios found for the cluster galaxies. Only LDR\,09 and LDR\,29
show roughly solar abundances, thus deviating significantly from the
cluster galaxies. We note that the galaxies identified in
Section~\ref{sec:ages} as having younger luminosity-weighted ages
(LDR\,08, 33, 34), are consistent with the cluster [Mg/Fe]--$\sigma$
relation.

\section{DISCUSSION}
\label{sec:discussion}
\subsection{The number and luminosity distribution of early-type
  galaxies in low-density environments}
\label{sec:disc_selection}
In the introduction to this paper we already hinted at the problems
concerned with the definition of a `field' galaxy sample. This
definition can range from selecting all galaxies, including systems in
rich clusters \cite{bens01}, to the very strict isolation criteria
applied in this paper and by Colbert et~al. (2001).  The definition
chosen necessarily depends on the specific application for which the
sample was designed.  Furthermore, observational or computational
restrictions may influence the sample construction.  Evidently,
therefore, when comparisons are made between different observations or
with theoretical model predictions one has to account carefully for any
differences in the definition of the samples. This caveat is especially
true when, as in this section, we discuss parameters related to the
number density of field E/S0s, or their luminosity distribution.

\begin{figure}
\psfig{file=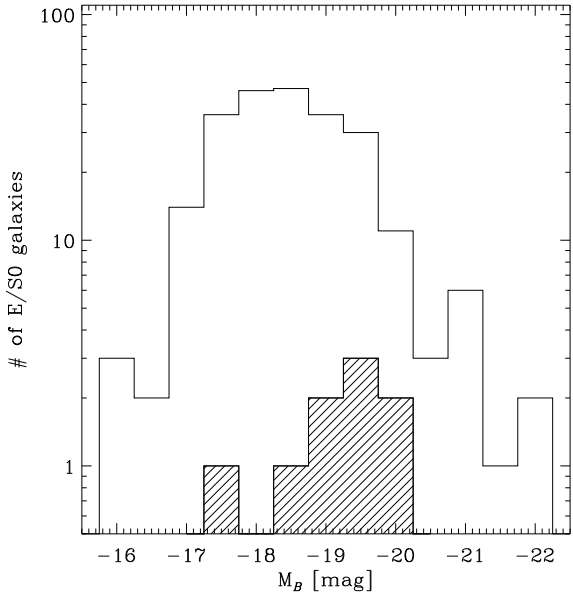,width=8.5cm}
\caption[]{\label{fig:lum_func}Luminosity function of all early-type
  galaxies in the FLASH survey for which we can reliably test our
  isolation criteria (see Section~\ref{sec:sample} for details). The
  hatched region shows our final sample of nine galaxies in low-density
  regions.}
\end{figure}

Do the LDR galaxies have the same luminosity function as early-type
galaxies in general? In Figure~\ref{fig:lum_func}, we show the
luminosity function of the putative parent sample, specifically those
FLASH survey galaxies with nominal types E or S0 lying within
7\,000~\kms. Overlaid as the hatched region is our sample of early-type
galaxies in low-density environments. Although we do not find any
galaxies in low-density environments brighter than $M_B \simeq -20$,
this is not surprising, since there are only few such galaxies in the
parent sample. Specifically, $\sim$95\% of the E/S0s in the parent
sample have $M_B > -20$. In a random nine-galaxy subsample, the
probability of selecting no galaxies above this luminosity threshold is
therefore $\sim{}0.95^9\approx0.6$. At low luminosities, the LDR
luminosity distribution seems to fall off much more rapidly than that
of the parent catalogue. This discrepancy may be caused by our counting
neighbours to a limiting apparent magnitude, rather than absolute
luminosity. This results in the nearby galaxies (and consequently the
low-luminosity galaxies) being subject to somewhat tighter selection
criteria than more distant (luminous) candidates. It is also possible
that many of the FLASH `early-type' galaxies in this luminosity range
are in fact spirals: Figure~\ref{fig:selection} shows that six out of
seven LDR candidate galaxies with $M_B>-18.25$ in fact exhibit spiral
morphologies in the CCD imaging.

The total number of galaxies of all morphologies, which satisfy our
isolation criteria, is 165. Thus, we estimate that the relative
fraction of {\em early-type}\/ galaxies, relative to all types, in
low-density environments is $0.08\pm0.03$. In this calculation we have
corrected for the ten unobserved candidates, assuming the same
confirmation rate as for the observed galaxies (37.5\%). This fraction,
which is of course sensitive to our specific selection criteria, could
be interpreted as the asymptotic limit of the morphology-density
relation (\eg\/ Dressler 1980, Dressler et~al.  1997, Hashimoto \&
Oemler 1999, Tran et~al. 2001). Note that even in the lowest-density
bin of the Dressler (1980) cluster study, the fraction of E/S0s is
still of order 50\%. At face value, then, the morphological mix in the
most isolated environments is dramatically different from that in the
outskirts of clusters.

Our sample selection criteria are similar in spirit to those of Colbert
et~al. (2001), who compiled an all-sky sample of 30 early-type galaxies
from the RC3 within 9\,900~\kms. Their isolation criteria are slightly
more stringent than ours, since they require that their galaxies have
no known neighbour within a radius of 1 $h^{-1}_{100}$~Mpc and
$\pm1000$~\kms. It is interesting that this sample has no galaxies
overlapping with ours. The Colbert et~al. sample in principle probes a
volume $\sim$150 times larger than the LDR sample, at least for the
brightest galaxies where both catalogues are volume limited. Accounting
for the different limiting magnitudes, we estimate that the galaxies in
the Colbert et~al. sample have a lower space density, by a factor of
$\sim$40, than those of our sample. This difference can probably be
explained by the stricter isolation criteria and by the brighter
magnitude limit of the RC3, but there could also be contributions from
incompleteness in the RC3 and available redshift data. The two samples
have very different luminosity distributions, with the Colbert et~al.
galaxies spanning $M_B= -19.6$~to~$-22.3$~mag, whereas our selection
criteria produce a sample with $M_B= -17.6$~to~$-20.1$~mag. As noted
above, the lack of very luminous galaxies in the LDR sample is not
inconsistent with the parent catalogue luminosity function. In the
Colbert et~al. sample, the brighter magnitude limit and much larger
survey volume yield a final sample with more highly-luminous systems.

Together, the LDR and Colbert et~al. samples indicate that there do
exist bona-fide early-type galaxies in low-density regions (selected by
criteria at least as stringent as ours), with luminosities brighter
than $M_B \sim -17.5$~mag. These galaxies are scarce in the nearby
universe, although quantitatively the number density of such galaxies
is clearly determined by the selection criteria imposed.

We note here that an alternative approach to defining a `field' sample of
early-type galaxies has been pursued by Grogin \& Geller (1999, 2000).
By smoothing the CfA2 redshift survey map on 5~\hmpc\/ scales, these
authors define large-scale ($\sim$30~\hmpc) `void' regions within the
survey, and select galaxies from these voids for follow-up. The
selection therefore differs from our method (and from that of Colbert
et~al.) in being less `local': a galaxy would not be excluded by the
presence of immediate neighbours, if its large-scale environment is of
low galaxy density.

Finally, we conclude that the selection criteria of Bernardi et~al.
(1998), who compile a sample of 631 `field' early-type galaxies (within
7\,000~\kms) must be very much less stringent than ours, and probably
include galaxies which reside in small groups or clusters. This is
probably also true for all studies at significant redshifts, where
redshift catalogues are often very much incomplete, or where galaxies
are selected only by projected distribution on the sky.

\subsection{Morphological merger signatures}
\label{sec:disc_morph}
The pioneering work of Schweizer and collaborators (\eg\/ Schweizer
et~al. 1990, Schweizer \& Seitzer 1992, see also Forbes 1991) has
linked the occurrence of morphological disturbances in early-type
galaxies such as ripples and shells, which are taken to be signs of
merger events (Hernquist \& Quinn 1988, 1989), to the presence of young
stellar populations.

A visual inspection of our optical images (Smith et~al., in
preparation) reveals that the LDR sample shows a variety of
morphological features indicative of recent merging and/or interaction
events. In particular, disturbed interacting companions are seen in
five cases (LDR\,08, 09, 19, 20, 33), tidal debris in four (LDR\,08,
09, 20, 33), and secondary intensity maxima along the major axis in two
(LDR\,20, 29)\footnote{These features could be interpreted as Lindblad
  or ultraharmonic resonances of an inner bar (see \eg\/ van den Bosch
  \& Emsellem 1998) or as the inner cut-off points of a `Freeman type
  II' disk \cite{free70}.}. There are prominently blue, and presumably
star-forming, circumnuclear rings in two galaxies (LDR\,08, 34).
Combining these indicators, six sample members show some peculiarity
(LDR\,08, 09, 20, 29, 33, 34). The remaining three galaxies (LDR\,14,
19, 22) seem to be comparatively undisturbed. Five galaxies (LDR\,08,
09, 14, 22, 33) are listed in the Arp \& Madore (1987) catalogue of
peculiar galaxies. For detailed comments on individual galaxies, see
Appendix~\ref{sec:ind_comments}.

Taking the detection rate of shells and tidal tails of 41\% from the
Colbert et~al. analysis, we should find $\sim$4 such galaxies in our
sample of nine, which is in good agreement with the above. The
fraction of galaxies in groups/clusters with shells or tidal tails is
much lower (\ie\/ 8\%, Colbert et~al. 2001). However, the hostile
cluster environment may destroy these morphological signs of merging
very quickly and it is therefore difficult to compare either the
intrinsic frequency of mergers, or when they happened.

We note that the lenticular galaxies in low-density regions have not
necessarily experienced the same formation history as cluster S0s,
despite their similar morphological appearance. In clusters it is
possible to create S0s via tidal stripping by intra-cluster gas, a
mechanism which does not affect galaxies in the field. In these
low-density environments, it is more likely that S0s form by
minor-mergers and/or accretion of gas which settles in a disk. LDR\,08
(see Appendix~\ref{sec:ind_comments}) may perhaps be an example of this
latter process in action. Due to infall from the field, clusters are
likely to contain galaxies of either formation type, or which have
experienced both of these pressures towards S0 morphology.

\subsection{Ongoing star-formation in LDR galaxies?}
\label{sec:disc_oii}
Compared to the cluster sample, the LDR sample shows both a higher
fraction of galaxies with emission (field: 56\%; cluster 29\%) and
slightly stronger emission lines. The average \oiiib\/ equivalent width,
for galaxies with significant emission, is $-0.88\pm0.17$~\AA\/ and
$-0.55\pm0.19$~\AA\/ for the LDR and cluster sample respectively.  This
`enhanced' emission in low-density regions is good evidence for a
prolonged, or at least more recent, period of star-formation activity.

This result can be set in respect not only to a local cluster
environment but also to higher redshifts. The \oii\/ emission strength
for our {\em nearby}\/ sample of LDR galaxies is in the range of 3 to 7
\AA\/ (equivalent width). This is much lower than Schade et~al. (1999)
found for their sample of field early-type galaxies at medium redshift
($0.2 \le z \le 1.0$), where one third of the ellipticals show \oii\/
emission in excess of 15~\AA. At face value, this indicates that the
rate of star-formation in E/S0s outside clusters has declined between
$z\simeq 0.5$ and the present day. The \oii\/ emission strengths for
LDR galaxies translate into star formation rates of $\le 0.14$ solar
masses per year \cite{ken92}, which is small compared to the medium
redshift sample.

Our observations probe only a nuclear region of each of the galaxies
and so any star-formation at larger radii will not be included in our
measurements. Although a good fraction of our galaxies show tails,
rings and debris at large radii, the surface brightness of these
features is very low. Only two galaxies (LDR\,08, 34) show prominent
circumnuclear rings in our U-band images, and the \oii\/ emission is
indeed stronger where the slit intersects these rings. This means that
`aperture corrections' for \oii\/ equivalent widths may potentially be
large, and highly unpredictable.

We conclude that our \oii\/ equivalent width measurements provide some
evidence for a reduction in star formation rate between $z\sim0.5$ and
the present time, but that this result may be sensitive to the
distribution of star formation activity within the galaxies.

\subsection{Luminosity-weighted age and metallicity distributions}
\label{sec:disc_age}
The high frequency of disturbed morphologies, the existence of
interacting neighbours and nebular emission lines in our low-density
region sample are all good evidence for merger induced star-formation
at low redshifts. Yet none of the galaxies currently exhibits a high
level of star-formation.

In Section~\ref{sec:ages}, we estimated luminosity-weighted ages and
metallicities for the LDR galaxies.  The luminosity-weighted ages range
from as old as the elliptical galaxies in the Fornax cluster
($\sim$10~Gyr) to as young as $\sim$3~Gyr, comparable with the merger
galaxy NGC\,1316 in Fornax. In the Fornax cluster, there is a clear
distinction between elliptical galaxies being all old, and the
lenticular galaxies showing a large spread in age.  This separation is
not seen so clearly in the LDR sample, as there are two ellipticals
with luminosity-weighted ages smaller than $\sim$8~Gyr. The overall age
distribution of the early-type galaxies in low-density regions is
similar to the distribution of the Fornax S0s, while the average age of
the LDR sample is 2--3~Gyr younger than the Fornax sample of early-type
galaxies.

Bernardi et~al. (1998) investigated the Mg$_2$--$\sigma$ relation in
their sample of field early-type galaxies drawn from the ENEAR survey
\cite{dcos00}. They detected a small but significant offset of
0.007~mag in the zero-point between the field and cluster/group
galaxies (the field having on average weaker Mg$_2$ absorption).  Using
stellar population models, they translated this difference into an
average luminosity-weighted age difference of $\sim$1~Gyr (the field
being younger). We discussed already the difference in selection
criteria, which complicates a direct comparison of these results with
ours.  However, we find no significant shift in the Mg$_2$--$\sigma$
zero-point, although we find a larger age difference between the
low-density environment and the cluster. This can be explained by the
young stars having also a higher metallicity, a so-called
age--metallicity anti-correlation \cite{coll99,tra00b}. This effect can
be seen directly in panel~c of Figure~\ref{fig:age_metal_MB} where we
demonstrate that the low-density region galaxies are on average more
metal rich than their cluster counterparts.  Anti-correlations between
the age, or the metallicity with the abundance ratios can also
contribute to the tightness of the Mg$_2$--$\sigma$ relation
\cite{tra00b,kun01}. However, since we do not find a significant
difference in the [Mg/Fe]--$\sigma$ relation between the cluster and
low-density environments, it seems unlikely that the [Mg/Fe] ratios
play a major role in stabilising the Mg--$\sigma$ relation.  We
conclude, that the analysis of the Mg$_2$--$\sigma$ relation can be
quite ambiguous and even mis-leading if an anti-correlation of age and
metallicity is present.

Trager et~al. (2000b) note that early-type galaxies, regardless of
their local environment, populate a two-dimensional plane in the
four-dimensional space of [M/H], $\log t$, $\log \sigma$, and [Mg/Fe] where,
at a given velocity dispersion, galaxies with younger ages have higher
metallicity. Considered jointly, our samples of galaxies in low-density
regions and in the Fornax cluster follow a similar relation. Only
LDR\,08 deviates significantly from the relation.

Although the sample is too small to permit a detailed comparison of
spectroscopic results with the presence of morphological peculiarities,
we note that the youngest galaxies in the LDR sample (LDR\,08, 33, 34)
all show clear signs of interaction, or show blue rings near the nucleus.
(see Section~\ref{sec:disc_morph}). The two oldest galaxies (LDR\,19, 22),
which also lie on the cluster [Fe/H]$-M_B$ relation, are among the three
most morphologically regular LDR galaxies.

\subsection{Confrontation with semi-analytic models}
\begin{figure*}
  \psfig{file=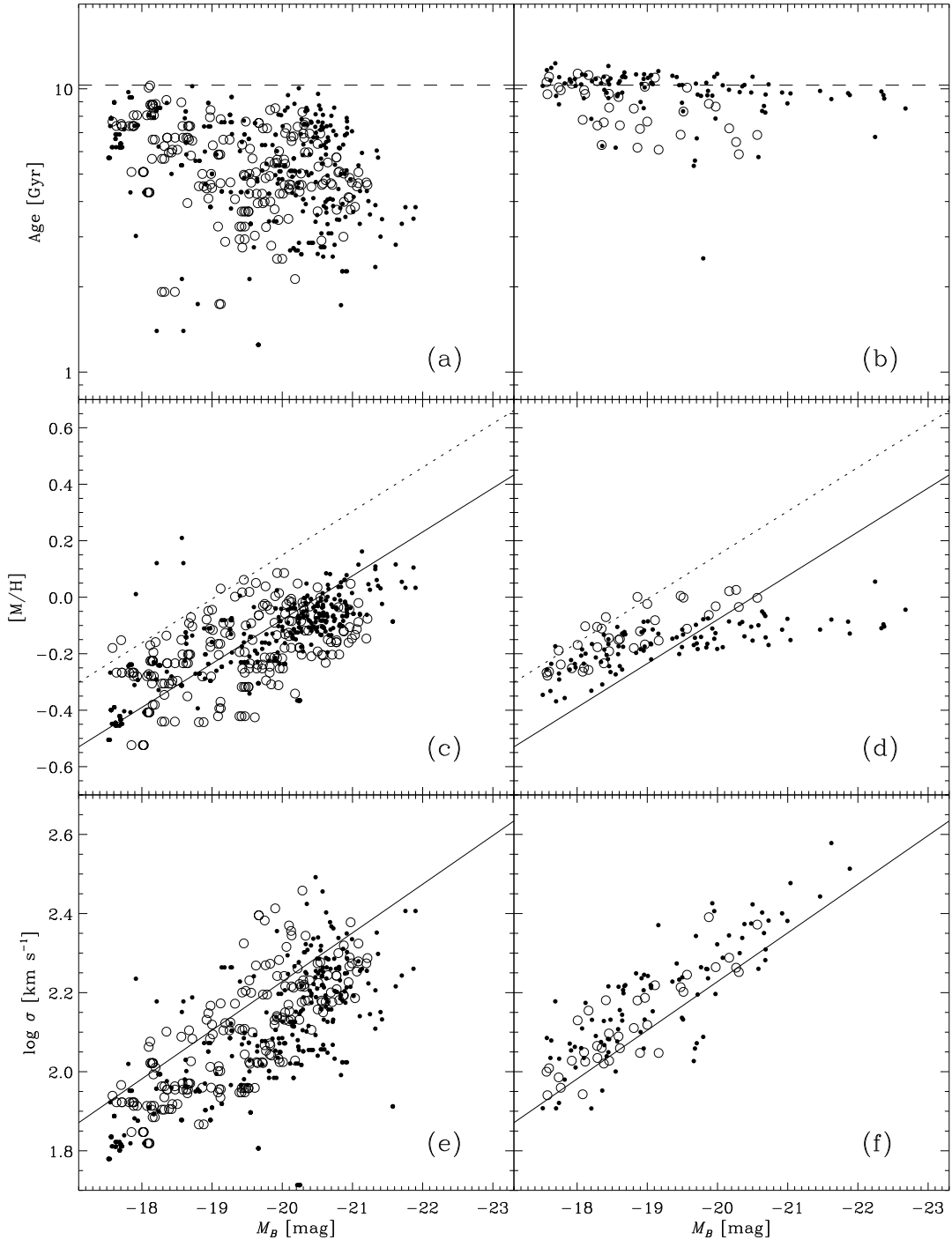,width=16.0cm}
\caption[]{\label{fig:cole}Predictions on the semi-analytic galaxy
  formation scenario \cite{bau96,col00}. Luminosity-weighted ages and
  metallicities are plotted as a function of $M_B$ for field galaxies
  (panels a, c) and cluster galaxies (panels b, d) with $M_B <
  -17.5$~mag. In the models the separation between low-density
  environments and clusters was defined by the dark matter halo mass,
  with clusters having dark matter halos of $>10^{14}$~M$_\odot$ and
  low-density environments having dark matter halos of
  $<10^{13}$~M$_\odot$.  Ellipticals (filled dots) are defined as
  galaxies with $0.6 < {\rmn B/T} \le 1.0$, whereas lenticular galaxies
  (open circles) have $0.4 < {\rmn B/T} \le 0.6$ (where the B/T ratio
  is derived from the predicted B-band luminosity). The central
  velocity dispersion was estimated by the rotation velocity of the
  bulge as given by the models divided by $\sqrt{3}$ (\ie\/
  $v_{bulge}/\sqrt{3}$). The linear relations indicated by solid,
  dashed and dotted lines are taken from Figure~\ref{fig:age_metal_MB}
  and can be used for a comparison of observed and predicted relations.
  See text for details.}
\end{figure*}

Recent semi-analytic models are able to predict the distribution of
luminosity-weighted age and metallicity in hierarchical galaxy
formation scenarios, as a function of halo mass. In this section we
compare our observational results with the models developed by the
`Durham' group \cite{bau96,col00}. We discuss the predictions of the
`Munich' models \cite{kau98} where these differ from the Durham
results. In Figure~\ref{fig:cole}, we show $V$-band weighted age,
metallicity and velocity dispersion\footnote{The velocity dispersion is
  inferred from the predicted circular velocity of the bulge, \ie\/
  $\sigma = v_{\rm bulge}/\sqrt{3}$} predictions for field and cluster
early-type galaxies, as a function of absolute blue magnitude $M_B$.
These predictions are from the so-called `reference' model of Cole et
al. (2000). The predicted luminosities, ages and metallicities of the
model galaxies are derived from ($V$-band weighted) superpositions of
individual simple stellar population models. Thus, to first order, the
parameters in Figure~\ref{fig:cole} should be directly comparable to
the estimated age and metallicities derived from our observations.
Only very recent bursts ($<$1~Gyr) could introduce effects which would
invalidate this comparison. In the models we have defined galaxies in
low-density environments as those residing in dark-matter halos with
mass $<10^{13}M_\odot$, while `cluster' members are located in halos of
mass $>10^{14}M_\odot$. The morphologies of the model galaxies are
assigned based on the bulge-to-total B-band light ratio: ellipticals
have 0.6 $<$ B/T $<$ 1.0, while lenticulars have 0.4 $<$ B/T $<$ 0.6.

\subsubsection{Ages}
The Durham models predict that ellipticals and lenticulars in clusters
should have a mean luminosity-weighted age of $9.3$~Gyr with a small
spread in age ($\pm1.7$~Gyr). Some of the cluster lenticular galaxies
scatter towards younger ages. By contrast, the field E/S0s show a more
evenly populated distribution of ages from 2--10~Gyr, and exhibit
younger ages on average (mean age: $5.5$~Gyr, see
Figure~\ref{fig:cdm_carlton} \& \ref{fig:cole}). The lower ages for
field E/S0s result from the greater probability of low-redshift mergers
where relative velocities are small (specifically, through the
dependence of the dynamical friction timescale on halo mass: Equation
4.16 of Cole et~al.). Younger ages are predicted for the more luminous
field galaxies, which are the last to form in these models.

Comparing our observed luminosity--age relations
(Figure~\ref{fig:age_metal_MB}a, b) with the semi-analytic predictions,
we find good agreement, at least qualitatively: the LDR sample exhibits
greater spread in age, and is offset to a mean luminosity-weighted age
significantly smaller than that of the Fornax elliptical galaxies.
While the Fornax sample does contain three lenticular galaxies
(labelled in Figure~\ref{fig:age_metal_MB}) with measured age less than
$\sim$8~Gyr, these represent only 18\% of the sample; for the LDR
galaxies, the equivalent proportion is 67\%.

Our data therefore support the Cole et~al. predictions in respect of
the luminosity-weighted age {\em vs}\/ luminosity relationships. Note
that the Munich models by contrast predict a much smaller age
difference between field and cluster (although in the same sense), and
a strong age--luminosity correlation in clusters which is not observed
in Fornax (Figure 4 of Kauffmann \& Charlot 1998).

\subsubsection{Metallicities}
For the metallicity distributions, the situation is more complex.  In
the Durham models, a luminosity--metallicity relation is predicted both
in the field and in clusters, but its shape, zero-point and scatter are
sensitive to the environment. In the model, field galaxies with $M_B >
-20$ have typically lower metallicities (at fixed luminosity) than in
clusters. The relation flattens in clusters, such that the field and
cluster metallicity distributions are similar for the brightest
galaxies (see Figure~\ref{fig:cole}).

Contrary to these predictions, our observational results
(Figure~\ref{fig:age_metal_MB}b, c) favour a higher metallicity (at
fixed luminosity) in the field, by a factor of $1.7\pm0.1$. All of our
LDR galaxies are fainter than $M_B \simeq -20$, where the models
predict an offset in the opposite sense. Moreover, the predicted
flattening of the cluster luminosity--metallicity relation at bright
magnitudes is not observed in the Fornax data.  The predicted
luminosity--metallicity curvature also generates a flattening of the
colour--magnitude relation for bright cluster ellipticals, which is not
observed (see Figure~13 of Cole et~al. 2000). In fact, the failure to
match the observed slope and form of the colour--magnitude relation has
been key difficulty in the development of semi-analytic galaxy
formation models (\eg\/ Kauffmann, White \& Guiderdoni 1993; Baugh
et~al. 1996). While an acceptable fit to the colour--magnitude data can
be obtained by adjusting parameters regulating yield and supernova
feedback (as in `Model A' of Kauffman \& Charlot 1998), this approach
leads to unreasonable predictions for other observables, such as the
luminosity function, and so seems not to be sustainable \cite{col00}.
We note that the Kauffman \& Charlot strong-feedback model predicts
that field and cluster galaxies follow identical
metallicity--luminosity relations, in contrast to our observations.

We may summarize that the metallicity predictions disagree with the
data on two counts: (i) the predicted metallicity is too low for the
most luminous galaxies in clusters, with respect to less-luminous
members; and (ii) the models predict a metallicity for the field
galaxies which is too low compared to cluster galaxies, especially at
lower luminosities. These two effects could result from the same
cause, if the models unduly suppress metal production in galaxies which
are dominant within their dark-matter halo. Field galaxies and,
progressively, the brighter members of clusters spend a greater
fraction of their histories in this privileged status. We speculate
that an excessive cooling of low-metallicity halo gas onto such
galaxies\footnote {\eg\/ through the models' neglect of energy
  injection from supernovae or AGN \cite{bow01}.} could be a cause of
the poor agreement with observed luminosity--metallicity relations.

There are alternative explanations for the field versus cluster
metallicity offset. In particular, there are several ways in which the
models do not predict exactly the same quantity as is measured here.
Firstly, the models predict total $V$-band weighted metallicity, where
our observations measure the metallicity only via the combination of
metal lines which contribute to the Fe5015 or [MgFe] indices; the
effects of non-solar abundance should be explicitly taken into account
for a fair comparison. More seriously perhaps, our data probes only the
central regions of the galaxies, while the semi-analytic models predict
`global' quantities for each galaxy. Hydrodynamical simulations of
galaxy mergers (\eg\/ Mihos \& Hernquist 1996) suggest that gas is
driven to the centre of the remnant; star-formation is then likely to
occur primarily in the nuclear region (perhaps in the central `blue
rings' observed in LDR\,08 \& LDR\,34). It is possible therefore, that
in galaxies which formed new stars during a recent merger, radial age
and metallicity gradients may be stronger than systems which have long
been undisturbed (\eg\/ in clusters). This effect would clearly act to
exaggerate the differences between the field and the cluster samples.

Our results can perhaps be described through a simplistic argument
based on the cosmological evolution of the average stellar metallicity.
If all merging and star formation ceases in cluster ellipticals after
some redshift $z_{\rm clus}$, then the average stellar metallicity at
that time [M/H]$(z=z_{\rm clus})$ will be `frozen' into the cluster
population. In the field, the stellar metallicity continues to rise
through the incorporation of processed material into new generations of
stars, a process which continues to $z=0$, so that the metallicity in
the field is [M/H]$(z=0)$. For an estimate of the rate of increase in
stellar metallicity, we may take a factor of 1.2 per unit $z$ (taken
from Figure 14 of Cole et~al.). Then using our measured metallicity
offset between field and cluster, we obtain the cluster `freezing'
redshift: $z_{\rm clus}=2.9\pm0.4$. This estimate is similar to the
formation redshifts implied by the homogeneity of the colour--magnitude
relation \cite{bow92b,kod98,ter01} and is consistent with an age of
9--11~Gyr for the cluster ellipticals, for reasonable cosmologies.

\subsubsection{Abundance ratios}
Thomas \& Kauffmann (1999, see also Thomas 1999) presented preliminary
results from their semi-analytic galaxy formation models (CDM universe)
for the distribution of [Mg/Fe] in galaxies as a function of
environment, morphology and luminosity. In these models, the most
luminous ellipticals are the last to form, incorporating, at least for
field galaxies, the ejecta of SN~Ia, as well as SN~II. Hence Thomas \&
Kauffmann find a trend that the [Mg/Fe] ratio decreases (to
approximately solar values) with increasing galaxy mass. This is even
more pronounced in low-density environments as the merging process and
hence star-formation continues to take place up to the present day.
Fainter galaxies are predicted to span a range in [Mg/Fe] in all
environments. In essence, the hierarchical galaxy formation scenario
predicts that present day early-type galaxies in low-density regions
show on average lower [Mg/Fe] abundance ratios than their brethren in
clusters.

As pointed out by Thomas \& Kauffmann, the observed trends in cluster
early-type galaxies are in stark contrast to the predictions of the
hierarchical galaxy formation model. We have extended the observations
to low-density regions and find an increasing trend of [Mg/Fe] with
central velocity dispersion, similar to that shown by cluster galaxies,
albeit with a larger scatter. Thus, our data further highlight the
failure of existing hierarchical models in explaining the observed
abundance ratios in E/S0s.

The super-solar values of [Mg/Fe] measured for the young galaxies in
our sample are surprising, because in the merger of two disk galaxies
with extended star-formation histories at low redshifts, one would
expect to produce an early-type galaxy with solar abundance ratios. A
possible explanation could be that the progenitor either harboured
stars which already match the stellar composition of a typical
early-type galaxy in clusters, or that the new stars, produced in a
rapid star-burst yielding super solar abundance ratios, dominate the
light of the galaxy as we see it today.

\section{CONCLUSIONS}
\label{sec:conclusion}
We have presented an analysis of a sample of nine nearby early-type
galaxies (3 elliptical, 6 lenticular), selected from a redshift survey
to reside in low-density regions. The sample is drawn from a sky-area
of $\sim$700 deg$^2$, to a redshift limit of 7\,000~\kms. Our stringent
selection criteria allow only up to two neighbours within a search
radius of 1.3~Mpc ($H_0 = 72$~\kmsM, q$_0 = 0.3$) and $\pm350$~\kms.
While the sample size is small, we emphasize that the multiple,
well-defined selection criteria guarantee that these galaxies are of
E/S0 morphology, and reside in large-scale environments of very low
density. We have investigated the Mg$_2$--$\sigma$ relation, and the
luminosity-weighted age, metallicity and abundance ratio distribution.
Our results have been compared to early-type galaxies in the Fornax
cluster and with the predictions for hierarchical galaxy formation.

The principal conclusions of our study are as follows:

\begin{itemize}
  
\item Elliptical and lenticular galaxies are rare in the `field', and
  account for only $\approx$8\% of the galaxy population in the
  lowest-density environments. The existing small samples of early-type
  galaxies in low-density environments give no evidence for any
  significant departure from the luminosity function of E/S0 galaxies
  in denser environments.
  
\item Five out of nine (56\%) sample galaxies show disturbed
  morphologies (\ie\/ tidal tails or debris, blue circumnuclear rings),
  which is interpreted as evidence of late merger events in these
  galaxies. Furthermore seven galaxies show close neighbours of which
  five exhibit signs of ongoing interaction with the main galaxy.
  
\item Compared to the cluster galaxies, our sample shows both a
  marginally higher incidence of \oiii\/ emission, and slightly
  stronger emission where present. However, we do not find galaxies
  with significant ongoing star-formation (\eg\/ \oii\/ $>10$~\AA).
  Relative to studies of early-type galaxies at intermediate-redshift,
  this indicates a significant decline in star-formation activity
  compared to the field population at a redshift of $z=0.5$.

\item The Mg--$\sigma$ relation in low-density regions is
  indistinguishable from that of cluster E/S0s. However, Mg--$\sigma$
  alone is a poor diagnostic tool for detecting differences in
  star-formation history: intrinsic (anti-)correlations between age,
  metallicity and abundance ratios have degenerate effects which can
  conspire to maintain the scatter and zero-point of the relation.
  
\item Early-type galaxies in low-density environments exhibit a broad
  distribution of luminosity-weighted ages, being on average younger
  than cluster ellipticals, while the distribution is more similar to
  that of lenticular galaxies in clusters. Taking the early-type galaxy
  population as a whole, the low-density regions harbour galaxies with
  2--3~Gyr younger luminosity-weighted ages than their brethren in
  clusters. This result is robust against the specific selection of the
  comparison sample since imposing an upper $\sigma$ limit does not
  change our result significantly while matching the luminosity
  distribution at the bright end increases the significance of our
  result. The younger ages of early-type galaxies in low-density
  regions is predicted by hierarchical galaxy formation, where the
  field population forms later and also experiences merger-induced
  star-formation episodes at lower redshifts.  We note that the
  youngest galaxies in our sample (LDR\,08, 33, 34) all show clear
  signs of interaction, or show blue rings near the nucleus, the latter
  being suggestive of gaseous accretion.
  
\item The luminosity-weighted metallicities of E/S0s in low-density
  environments are larger than for cluster members of similar
  luminosity. This effect ($\Delta$[Fe/H]~$\approx$0.2~dex) is not
  seen in semi-analytic models, which predict an offset in the opposite
  sense, at least for the luminosity range probed by our data.
  Furthermore, the current generation of semi-analytic models cannot
  explain the observed mass-metallicity relation of bright cluster
  members without compromising the reproduction of other observables,
  such as the luminosity function. These disagreements between the
  observations and semi-analytic model predictions highlight important
  shortcomings in the detailed treatment of the star-formation
  processes in present models.

\item The E/S0 galaxies in our low-density sample exhibit mostly
  super-solar [Mg/Fe] ratios. The the non-solar [Mg/Fe] values rise
  with central velocity dispersion, following a trend similar to that
  of cluster members. By contrast, hierarchical galaxy formation models
  predict approximately solar abundance ratios, at least for the
  brightest galaxies in low-density regions.

\end{itemize}

In summary, this study, though for a small sample, finds results
consistent with one of the central predictions of hierarchical galaxy
formation models: the formation of early-type galaxies continues to
$z\la1$ in low-density environments, while those in clusters formed
most of their stars at $z\ga2$. On the other hand, our results
underscore the models' present failure to reproduce the observed
luminosity--metallicity trends and their apparent dependence on
environment. A future generation of models must also overcome the stiff
challenge of generating super-solar [Mg/Fe] ratios, even in galaxies
formed by late-time merging of potentially spiral galaxies with
extended star-formation histories.

\section*{ACKNOWLEDGEMENTS}
This research has made use of the NASA/IPAC Extragalactic Database
(NED) which is operated by the Jet Propulsion Laboratory, California
Institute of Technology, under contract with the National Aeronautics
and Space Administration. The Cerro Tololo Inter-American Observatory
(CTIO) is operated by the Association of Universities for Research in
Astronomy, Inc. under a cooperative agreement with the National Science
Foundation. {\sc Iraf} is distributed by the National Optical Astronomy
Observatories which is operated by the Association of Universities for
Research in Astronomy, Inc. under contract with the National Science
Foundation. HK acknowledges support from the PPARC grant `Extragalactic
Astronomy \& Cosmology at Durham 1998-2000' and the ESO fellowship
program. RJS acknowledges partial financial support from
FONDECyT--Chile (Proyecto 3990025). RLD gratefully acknowledges the
award of a Research Fellowship from the Leverhulme trust and a PPARC
Senior Fellowship. We thank Carlton Baugh, Shaun Cole and
collaborators, for providing access to their semi-analytic model
predictions, and for helpful discussions concerning the interpretation
of the models.

\appendix
\section{Aperture corrections for Fornax galaxies}
\label{sec:aperture}
In this section we describe briefly the aperture corrections applied to
the Fornax cluster galaxies in order to match our nominal aperture of
$4 \arcsec \times 2 \arcsec$ at 5\,000~\kms. This is equivalent to
1.08~kpc diameter (H$_0 = 72$ \kmsM, q$_0 = 0.3$).

At the distance of Fornax ($m-M=31.52\pm0.04$, Tonry et~al 2001) the
1.08~kpc diameter is equivalent to 11\farcs2. Extracting an equivalent
aperture from our slit observations (2\farcs3 slit width) yields
prohibitively long extraction. In order to overcome this problem we
extract an aperture of $2\farcs3 \times10 \arcsec$ (equivalent to
5\farcs5 diameter aperture) for all Fornax galaxies and correct the
remaining aperture difference by using the line strengths gradients for
individual galaxies (see Kuntschner 1998 for line-strengths gradients).

Here we first transformed the ``atomic'' indices measured in \AA\/ onto
a magnitude scale like the ``molecular'' index Mg$_2$. The conversion
between an index measured in \AA\/ and magnitudes is

\begin{equation}
  \label{equ:mag}
  {\rmn index\, [mag]} = -2.5 \log \left( 1 - \frac{\rmn index\, [\AA]}{\Delta \lambda}\right)
\end{equation}
where $\Delta \lambda$ is the width of the index bandpass (see \eg\/
WO97 and Trager et~al. 1998 for a list of bandpass definitions).
Following J{\o}rgensen (1997) we then correct the indices for aperture
effects in the following way:

\begin{equation}
  \label{equ:indexp}
  {\rmn index}_{norm}\,{\rmn [mag]} = {\rmn index}_{ap}\, {\rmn [mag]} - \alpha \log \frac{r_{ap}}{r_{norm}}
\end{equation}
where $\alpha$ is the slope of the radial gradient
$\{\Delta$~index~[mag] $/ \Delta \log r\}$. For the Fornax cluster
$\log (r_{ap}/r_{norm})=-0.31$.  After the correction the ``atomic''
indices are converted back to \AA\/ scale. This somewhat complicated
procedure was performed because of two reasons: (i) the gradients for
the Fornax galaxies where available only for indices on the magnitude
scale, and (ii) indices such as \HgF\/ which can show values close to
zero, and indeed negative values, are not suitable for a multiplicative
correction such as successfully used by J{\o}rgensen (1997) for other
atomic indices.

We emphasize that for each galaxy we have used the individually
determined line-strength gradients in order to provide the aperture
corrections. However, for the reader's guidance and future reference we
list in Table~\ref{tab:grad} the average gradients of {\em
  elliptical}\/ galaxies in our sample for a set of key line-strength
indices. The gradients are derived from data covering the range up to
$\sim$1.5 effective radii.

\begin{table}
  \caption[]{Average line-strength gradients for elliptical galaxies}
  \label{tab:grad}
  \begin{tabular}{lc} \hline
    index     & $\alpha = \frac{\Delta index [mag]}{\Delta \log r}$ \\ \hline
    G4300     & $-0.028\pm0.011$ \\
    Fe4383    & $-0.031\pm0.015$ \\
    C$_2$4668 & $-0.032\pm0.004$ \\
    \Hb       & $+0.000\pm0.006$ \\
    Fe5015    & $-0.019\pm0.009$ \\
    Mg$_1$    & $-0.040\pm0.015$ \\
    Mg$_2$    & $-0.066\pm0.019$ \\
    \mgb      & $-0.045\pm0.013$ \\
    Fe5270    & $-0.019\pm0.008$ \\
    Fe5335    & $-0.019\pm0.008$ \\
    Fe5406    & $-0.018\pm0.005$ \\
    \HgA      & $+0.034\pm0.011$ \\
    \HgF      & $+0.033\pm0.013$ \\ \hline
  \end{tabular}

    \medskip
    \begin{minipage}{6.0cm}
      Note: The gradients are given for all indices measured in
      magnitudes (see Equation~\ref{equ:mag}) and as a function of
      $\log r$.
    \end{minipage}
\end{table}

The gradients given here for Mg$_1$ and Mg$_2$ are similar to the one
used by J{\o}rgensen (1997, $\alpha = 0.04$) for both indices. The
aperture corrections of the velocity dispersions were performed using
the following formula:

\begin{equation}
  \label{equ:mgbp}
  \log \sigma_{norm} = \log \sigma_{ap} - \alpha \log \frac{r_{ap}}{r_{norm}}
\end{equation}
where $\alpha = -0.04$. This is the same correction strength as used by
J{\o}rgensen, Franx \& Kj{\ae}rgaard (1995).

\section{New stellar population models}
\label{sec:vazdekis}
In order to make age and metallicity estimates, we use the Vazdekis
(1999) models, which utilize the empirical stellar library of Jones
(1999) to predict line-strengths for a single-burst stellar population
as a function of age and metallicity. These models have recently been
updated (Vazdekis 2002, in preparation) incorporating the new
isochrones of Girardi et~al. (2000) and extensive empirical photometric
libraries, such as Alonso et~al. (1999). For some more detailed
descriptions see also Blakeslee, Vazdekis \& Ajhar (2001). The model
predicts, among other parameters, spectral energy distributions (SEDs)
in two regions in the optical wavelength range (3856-4476~\AA\/ and
4795-5465~\AA) for simple stellar populations (SSPs), with a range of
metallicities ($-0.7\leq \log(Z/Z_\odot) \leq+0.2$) and ages (1 to
17~Gyr). We use a Salpeter IMF (corresponding to the {\tt unimodal}
option with $\mu = 1.3$ in the Vazdekis models).  The spectral
resolution of the models is 1.8~\AA\/ (FWHM). If model predictions of
lower spectral resolutions are required, then the spectra can be simply
broadened. For this paper we smoothed the spectra to the spectral
resolution given by our data, \ie\/ 4.1~\AA.  Predictions for
line-strength indices as a function of age and metallicity are easily
obtained by measuring the indices directly on the predicted SEDs.
Previous models (\eg\/ Worthey 1994; Vazdekis et~al. 1996) used mostly
the Lick polynomial fitting functions \cite{wor94b,wor97} to relate the
strengths of selected absorption features to stellar atmospheric
parameters. The fitting functions are based on the Lick/IDS stellar
library (FWHM $\sim$9~\AA, Worthey et~al.  1994), this limits the
application of the models to strong features.

The Lick stellar library has not been flux calibrated, so offsets had
to be applied to data obtained with different spectrographs
\cite{wor97}. The new Vazdekis models provide flux calibrated spectra,
hence offsets should be small or zero.

The interpretation of our line-strength measurements for the field
early-type galaxies depends sensitively on the model predictions and
their accuracy. Since we are here using the predictions of the new
Vazdekis models which will be described in a later paper (Vazdekis
2002, in preparation) we present in the following paragraphs a short
comparison between the new models and the well established stellar
population predictions by Worthey (1994). For the comparison we have
chosen to use the \Hb\/ {\em vs}\/ [MgFe] diagram \footnote{${\rmn
    [MgFe]} = \sqrt{{\rmn Mg}\,b \times ({\rmn Fe5270} + {\rmn
      Fe5335})/2}$\,, \cite{gon93}}, since it is one of the key diagrams
in this paper.

\begin{figure*}
\psfig{file=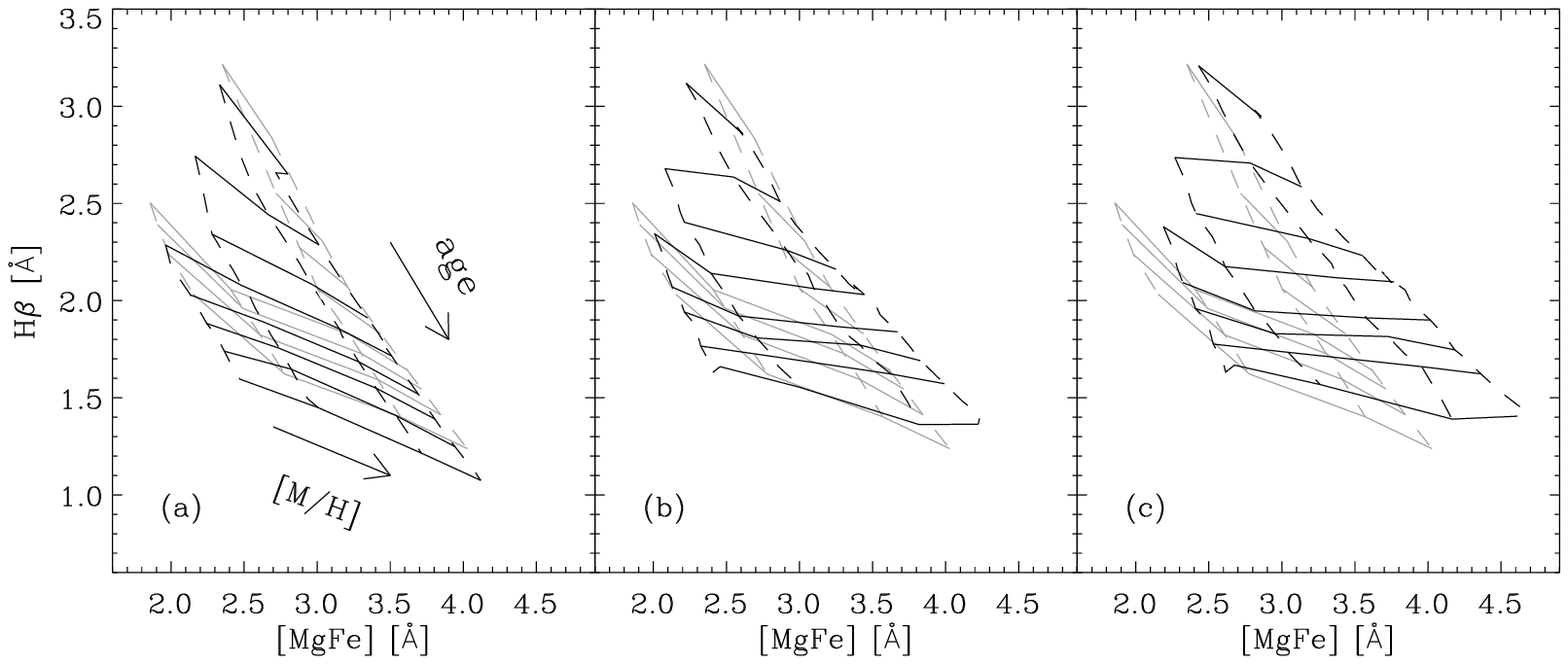,width=14.5cm}
\psfig{file=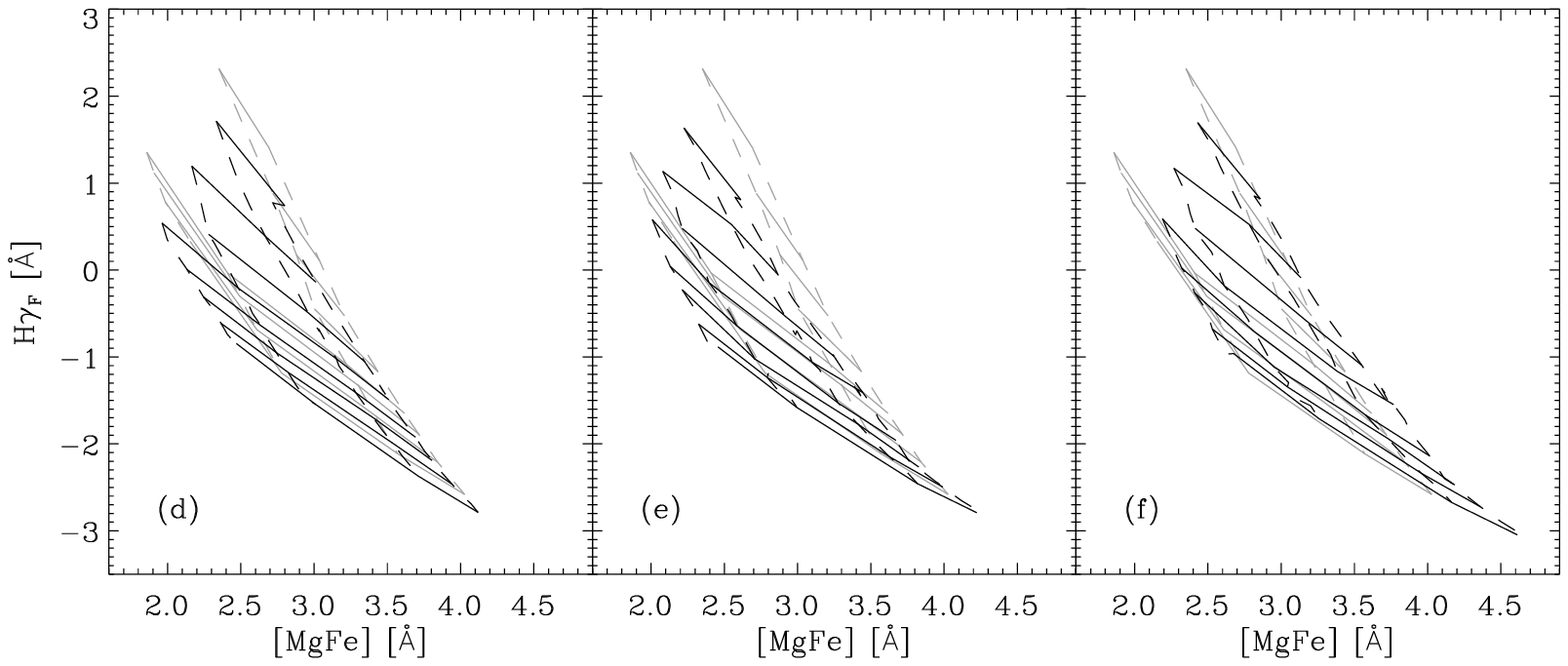,width=14.5cm}
\psfig{file=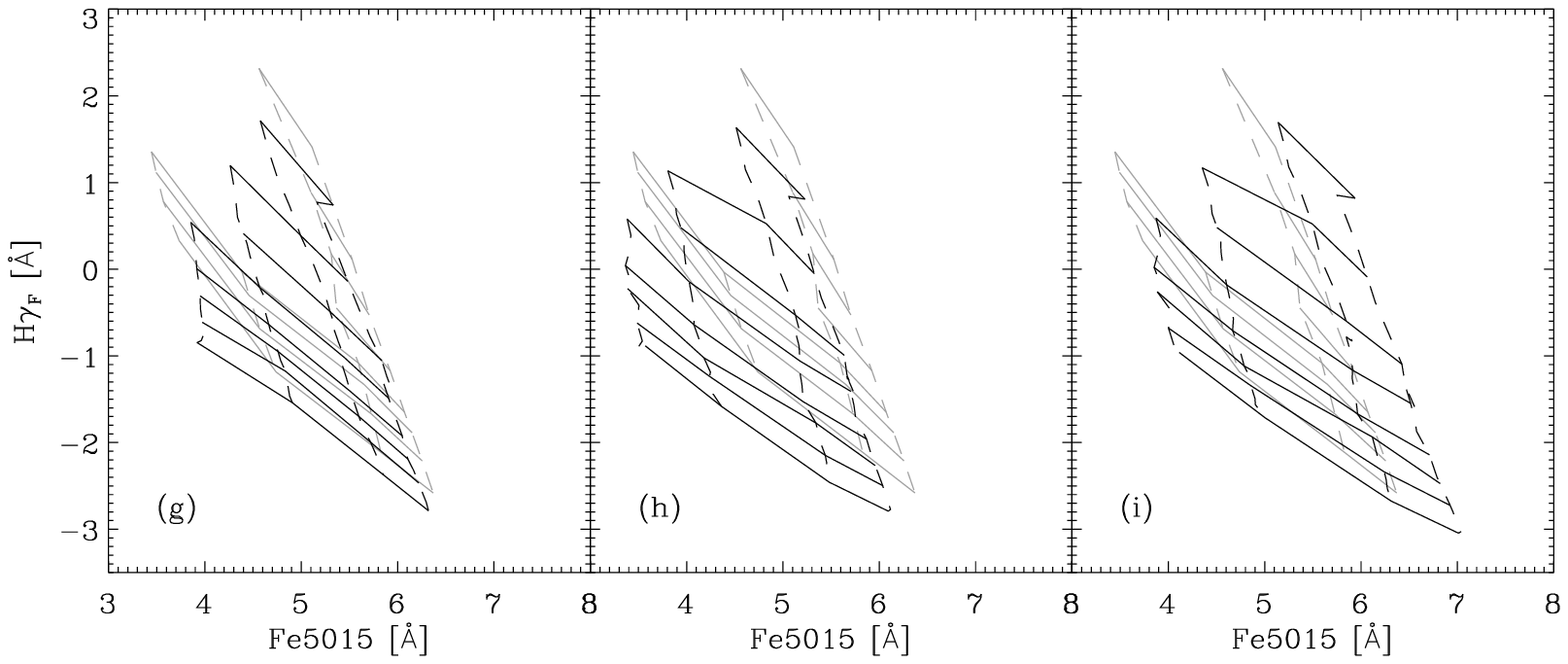,width=14.5cm}
\caption[]{\label{fig:models}Comparison of stellar population
  models in the \Hb\/ {\em vs}\/ [MgFe] diagram, the \HgF\/ {\em vs}\/
  [MgFe] diagram and the \HgF\/ {\em vs}\/ Fe5015 diagram (from top to
  bottom). The grey lines represent the models of Worthey (1994) and
  are given in all panels as a reference. The black lines represent
  three different versions of Vazdekis et~al. (1996) and Vazdekis
  (1999) models shown from left to right. The solid lines are lines of
  constant age, whereas the dashed lines are lines of constant
  metallicity. Both models are shown in age steps of 1.6, 2.5, 3.6,
  5.6, 8.0, 10.0, 12.6, 17.8~Gyr and metallicity steps of
  $-0.7$,$-0.4$, 0.0, 0.2 in log units (Salpeter IMF). The direction of
  increasing age and metallicity is indicated in panel (a). The Worthey
  models were interpolated to the given age and metallicity steps with
  the use of Dr. Guy Worthey's Home page ({\tt
    http://astro.sau.edu/worthey/}; version as of Jan 2001).
  Panels~(a), (d) \& (g): Vazdekis et~al. (1996) models using the
  Lick/IDS fitting functions. Panel~(b), (e) \& (h): Vazdekis (1999)
  models using the Jones (1999) library broadened to the Lick/IDS
  spectral resolution (\ie\/ $\sim$9~\AA). Panels~(c), (h) \& (i):
  Vazdekis (1999) models using the Jones library at 4.1~\AA\/ (FWHM)
  spectral resolution. See text for details.}
\end{figure*}

Firstly, we compare the Worthey (1994) models with a version of the
Vazdekis models \cite{vaz96} which also uses the Lick/IDS fitting
functions to predict index strength but has otherwise its own stellar
population prescriptions (Figure~\ref{fig:models}a). Overall the models
agree very well in this parameter space. The most important difference
is seen in the strength of \Hb\/ which shows a systematic offset
($\sim$0.15~\AA) towards smaller values at a given age in the Vazdekis
models. This shift is most likely caused by differences in the
isochrone temperatures (see Vazdekis et~al. 1996). We also note that
for ages larger than 8~Gyr and the lowest metallicity step (${\rmn
  [M/H]} = -0.7$) the models show systematically larger differences.
This is caused by the uncertainties involved in the treatment of the
horizontal branch. In summary the model predictions agree well when the
Lick/IDS fitting functions are used to calculate the line-strength
indices. Only small offsets in the absolute age/metallicity predictions
for a given index strength are present.

Secondly, we use the Vazdekis models to predict line-strength indices
for a given age and metallicity in combination with the new spectral
library by Jones (1999). In model terms this is a replacement of the
Lick/IDS fitting functions. Other stellar population parameters are
untouched. We emphasize, however, that we broadened the SEDs produced
by the Vazdekis models to the Lick/IDS spectral resolution \cite{wor97}
and measured the Lick/IDS indices directly on the spectra. The models
are shown in Figure~\ref{fig:models}b. Overall the models occupy the
same parameter space in the diagram and agree well on a general level.
In detail, however, the new models show a better separation of age and
metallicity effects in these coordinates than is indicated by the
Worthey (1994) models using the Lick/IDS fitting functions. In other
words, the lines of constant age (black solid lines) are almost
horizontal.  Furthermore, the new models predict for a given strength
of [MgFe] lower metallicities compared to Worthey (1994). This effect
increases towards greater ages and higher metallicities.

Although the differences in Figure~\ref{fig:models}b appear to be small
we note that they can change the predicted ages and metallicities for a
given line-strength significantly. For example, the predicted
line-strength for solar metallicity and an age of 12.6~Gyr in the
Vazdekis models corresponds to an age of 9~Gyr and ${\rmn [M/H]} =
+0.2$ in the Worthey (1994) models.

In the last diagram (Figure~\ref{fig:models}c) we show the new Vazdekis
models, using the Jones (1999) spectral library, at a resolution of
4.1~\AA\/ (FWHM). Here, the overall shape of the models does not change
very much compared to Figure~\ref{fig:models}b, but the whole model
grid is shifted towards higher values of [MgFe]. This is mostly a
reflection of the spectral resolution sensitivity of the Fe lines which
contribute to the [MgFe] index. This last diagram demonstrates clearly
the importance of using the right models when working at a higher
spectral resolution.

In this paper we use two other important age-metallicity diagnostic
diagrams which are presented in the following. Figures
\ref{fig:models}d-f and Figures \ref{fig:models}g-i show the model
comparisons for the \HgF\/ {\em vs}\/ [MgFe] and the \HgF\/ {\em vs}\/
Fe5015 diagrams, respectively. The differences between the models in
the \HgF\/ {\em vs}\/ [MgFe] are very small and do not lead to
significantly different conclusions about the luminosity-weighted ages
and metallicities. In the \HgF\/ {\em vs}\/ Fe5015 diagram we find good
agreement when the Lick/IDS fitting functions are used. However, when
we use the Vazdekis models with the Jones (1999) library broadened to
the Lick/IDS resolution, the predictions for the Fe5015 index differ
significantly. At a given strength of Fe5015 the Vazdekis models
predict higher metallicities compared to Worthey (1994). This effect
slightly increases for larger ages.

In summary we conclude that the relative ordering of an age or
metallicity sequence is mostly preserved whatever stellar population
model is used. The absolute ages and metallicities vary substantially
and remain insecure. The most significant differences in the
line-strength predictions between the Worthey (1994) and Vazdekis
(1999) models are caused by the change from the Lick/IDS fitting
functions to the Jones (1999) library and/or by measuring the indices
directly on the spectra rather than creating new fitting functions. The
differences in the stellar evolution prescriptions seem to have only
minor effects in the considered metallicity and age range.

We emphasise that the model predictions are based on the assumption of
a single metallicity and single age stellar population. Real galaxy
spectra most likely show a range in these parameters, due to stellar
population gradients and projection effects (\eg\/ Maraston \& Thomas
2000). Furthermore, none of the models treats the effects of non-solar
abundance ratios which undoubtedly can influence the predictions for
line-strength indices and therefore the resulting age and metallicity
estimates. We are also measuring luminosity-weighted properties, which
means that the presence of a young (\ie\/ bright) stellar component
will disproportionally affect the mean age and therefore will be
detected more easily.

\section{Comments on individual galaxies}
\label{sec:ind_comments}
Below we collate some miscellaneous notes on individual galaxies.
Where available we list other names from the NGC catalogue
\cite{nil73}, the ESO/Uppsala survey \cite{lau82} or the Arp \& Madore
(1987, hereafter AM87) catalogue of peculiar galaxies. Projected
distances are calculated for H$_0 = 72$~\kmsM\/ (q$_0 = 0.3$), assuming
homogeneous Hubble flow.

{\bf LDR\,08}\/ -- (ESO\,503-G005, AM\,1112-272): Listed in AM87 under
code 15 (Galaxies with tails, loops of material or debris) with comment
``E + faint irregular ring''. Our imaging shows that the galaxy
presents an elliptical-like core, but has two prominent irregularities:
an outer system of tidal tails (extending to 60\arcsec, $\sim$16~kpc
radius) and a slightly elliptical inner `ring', at a radius of
$\sim$5\arcsec ($\sim$1.2~kpc). The brightest part of the tidal tail is
conspicuously blue, and has a knotty structure, suggesting ongoing
accretion of a gas-rich, star-forming companion. This galaxy is
detected in H\,{\small I} by the `The H\,{\small I} Parkes All-Sky
Survey' (HIPASS, Barnes et~al. 2001).

{\bf LDR\,09}\/ -- (NGC\,3617, ESO\,503-G012, AM\,1115-255): Listed in
AM87 under code 23 (Close pairs, not visibly interacting) with comment
``Close pair (\#2: LSB spiral)''. LDR\,09 has boxy outer isophotes.
The companion (ESO\,503-G011) is located at a distance of $\simeq
5^m50^s$ (48~kpc projected distance), the redshift is
$2047\pm4$~\kms\/, $b_J=15.58$, morphology: Sa-b. Our imaging shows a
faint companion closer in (at 90\arcsec = 13~kpc), this object is
clearly interacting with the galaxy, producing a narrow linear tidal
feature, easily traced over $\sim$3.5~kpc. The satellite itself is of
unusual appearance, apparently having three distinct nuclei.

{\bf LDR\,14}\/ -- (ESO\,379-G026, AM\,1203-354): Listed in AM87 under
code 23, (Close pairs, not visibly interacting) with comment ``Close
pair (E + spiral)''. The companion, a spiral galaxy, is located at a
distance of $\simeq1^m25^s$ (21~kpc projected distance; J2000
$12^h06^m18\farcs0$, $-36\degr00^m17^s$). The redshift is unknown,
$b_J=15.99$. No clear signs of interaction.

{\bf LDR\,19}\/ -- (ESO\,442-G006) S0 with no significant peculiarities.
There is a late-type spiral companion $1^m40^s$ to the west. There
are two faint galaxies embedded in the outer isophotes, of which one shows
signs of tidal interaction.

{\bf LDR\,20}\/ -- (ESO\,381-G004) Our images show that this galaxy is
clearly undergoing interaction with a companion, which may have passed
through the centre of the galaxy. The companion, an elongated arc-like
structure (J2000 $12^h39^m07\farcs2$, $-34\degr45^m37^s$), at a
distance of $\simeq 1^m20^s$ (22~kpc projected distance) is of similar
colour to the dominant galaxy. Several broad streams of tidal debris
are visible. The galaxy shows symmetric secondary intensity maxima
along the major axis (see also LDR\,29).

{\bf LDR\,22}\/ -- (ESO\,382-G016, AM\,1310-362): Listed in AM87 under
code 8 (Galaxies with apparent companion(s)) \& 14 (Galaxies with
prominent or unusual dust absorption) with comment ``E with peculiar
linear companion + dust''. The S0 companion (ESO\,382-G017) is located
just outside our images, at a distance of $\simeq 3^m00^s$ (38~kpc
projected distance), and has $cz=3280\pm36$~\kms\/, $b_J=15.05$.  In
contrast to the AM87 study we do not find evidence for significant dust
in this galaxy.

{\bf LDR\,29}\/ -- (ESO\,445-G056) Possible cluster member, but very
large relative velocity. The galaxy is highly elongated, with
ellipticity 0.6--0.7 in the outer regions. The isophote shapes are
disky, in part due to two symmetric intensity peaks similar to those
seen in LDR\,20, which could be interpreted as Lindblad or
ultraharmonic resonances (see \eg\/ van den Bosch \& Emsellem 1998), or
as the inner cut-off points of a `Freeman type II' disk \cite{free70}.
There are no signs of ongoing interaction.

{\bf LDR\,33}\/ -- (AM\,1402-285): Listed in AM87 under code 2
(Interacting doubles (galaxies of comparable size)) with comment
``Interacting double''. The companion (J2000; $14^h05^m24\farcs7$,
$-29\degr07^m23^s$) is located at a distance of $\simeq 1^m10^s$
(27~kpc projected distance). The redshift and $b_J$ magnitude are
unknown. LDR\,33 appears slightly asymmetric, and seems to have broad
tidal streams similar to (but much weaker than) those in LDR\,20. The
companion (spiral?), with a dust-lane, shows evidence of tidal
stripping.

{\bf LDR\,34}\/ -- (ESO\,446-G049) The galaxy has a narrow blue ring
(radius 10\arcsec, 2.5~kpc) like that seen in LDR\,08 and a peculiar
outer disk/halo. No close, bright companions.

\bsp

\label{lastpage}

\end{document}